\documentclass[journal,twocolumn]{IEEEtran}

\usepackage{graphicx}

\ifCLASSINFOpdf
    \DeclareGraphicsExtensions{.pdf,.png,.jpg,.jpeg,.eps}
\else
    \DeclareGraphicsExtensions{.eps}
\fi

\usepackage{cite}

\usepackage[caption=false,font=footnotesize]{subfig}

\usepackage{amsmath}
\usepackage{amsfonts}
\usepackage{amssymb}
\usepackage{amsthm}
\usepackage{bm}
\usepackage{bbm}
\usepackage{dsfont}
\usepackage{mathalfa}
\usepackage[thinc]{esdiff}

\usepackage{algorithm}
\usepackage{algpseudocode}

\usepackage[normalem]{ulem}
\usepackage{ragged2e}
\usepackage{xcolor}
\usepackage[most]{tcolorbox}

\newcommand{\mb}[1]{\mathbf{#1}}
\newcommand{\mr}[1]{\mathrm{#1}}

\newcommand{\e}{\mathrm{e}}
\newcommand{\ji}{\mathrm{j}}

\newcommand{\E}{\mathbb{E}}

\begin{document}

\title{Data-Aided Asynchronous OFDM Integrated Sensing and Communications: A Mean-Field Variational Bayes Approach}

\author{
    \IEEEauthorblockN{
        Van-Chung Luu\IEEEauthorrefmark{1}\IEEEauthorrefmark{2},  
        Nuria González-Prelcic\IEEEauthorrefmark{2}, 
        and Duy H. N. Nguyen\IEEEauthorrefmark{1}
    }\\
    \IEEEauthorblockA{
    \begin{tabular}{l}
        \IEEEauthorrefmark{1}Department of Electrical and Computer Engineering, San Diego State University, San Diego, CA, USA\\
        \IEEEauthorrefmark{2}Department of Electrical and Computer Engineering, University of California, San Diego, CA, USA
    \end{tabular}
}}

\maketitle
\begin{abstract}
Integrated sensing and communication (ISAC) is regarded as a key technology for sixth-generation wireless networks, allowing sensing and communication operations to jointly utilize the same spectrum and hardware infrastructure. However, in practical uplink ISAC systems, timing offset (TO) and carrier-frequency offset (CFO) introduce phase distortions across subcarriers and OFDM symbols, which can severely degrade both data detection and sensing-parameter estimation. In this paper, we propose a data-aided variational Bayesian (VB) framework for asynchronous uplink OFDM-ISAC systems. Specifically, the received signal is modeled as a sparse multipath superposition, where the transmitted data symbols, complex path gains, spatial frequencies, delay-Doppler parameters, and synchronization parameters are jointly inferred. To enable tractable inference, we develop a mean-field VB algorithm in which von Mises distributions are used for gridless updates of the angular and delay-Doppler phase parameters, while a Gamma-Gaussian prior is adopted to promote path sparsity. A key feature of the proposed framework is its data-aided sensing capability: after initial 
pilot-based estimation, the detected data symbols are exploited as additional observations to refine the channel and sensing parameters. This substantially increases the effective sensing resources without requiring extra pilot overhead. The simulation results demonstrate that the proposed approach achieves superior performance compared with SAGE,  SBL, AB2FM, and pilot-only VB baselines in terms of symbol error rate, channel reconstruction accuracy, path-parameter estimation, TO/CFO estimation, and 3D 
localization accuracy. The results also demonstrate that ignoring TO and CFO leads to severe sensing degradation, highlighting the importance of synchronization-aware and data-aided receiver design for ISAC systems.
\end{abstract}

\section{Introduction}

Integrated Sensing and Communications (ISAC) is widely regarded as a cornerstone technology for future 6G wireless networks. Depending on the primary objective, these systems are generally categorized into communication-centric or sensing-centric frameworks \cite{zhang2020perceptive, cui2021integrating}. The communication-centric approach is particularly compelling, as it enables the reuse of existing infrastructure, hardware, and spectral resources for dual-functional tasks, thereby maximizing overall system efficiency \cite{gonzalez2024integrated}. Achieving this integration in practical wireless networks requires a waveform that can accommodate both functionalities within a common transmission structure. In this context, orthogonal frequency division multiplexing (OFDM) has emerged as one of the most practical waveform foundations for ISAC, owing to its widespread adoption in modern wireless systems and its flexible time-frequency structure, which enables sensing functions to be embedded into communication frames with relatively small architectural changes \cite{barneto2019full,11358863,nguyen2023multiuser,zhang2024crossdomain,he2024dualfunctional}. Beyond early studies that allocated subcarriers or power between communication and sensing \cite{cheng2021hybrid,wang2019power,shi2019joint,ahmed2019ofdm}, more recent OFDM-ISAC research has advanced toward cross-domain waveform optimization, dual-functional precoding, and sensing-oriented resource allocation strategies that jointly balance communication rate, sensing resolution, sidelobe suppression, and robustness \cite{zhang2024crossdomain,he2024dualfunctional,mura2025optimized,li2025sensing}. The multicarrier structure of OFDM makes it well suited for wideband transmission in frequency-selective channels. This feature is particularly beneficial in high-frequency bands, such as the Ku-band, where large bandwidths can be leveraged to improve both communication throughput and sensing resolution \cite{gonzalez2025six}.

While promising in theory, the practical realization of ISAC gains is hindered by significant hardware and environmental constraints. Specifically, ISAC systems are subject to several practical impairments, such as synchronization mismatch, mobility, and multipath propagation \cite{wu2024sensing, 10207823, ni2021uplink, luo2024integrated}. In high-frequency and wideband scenarios, these impairments become even more critical, as synchronization mismatch introduces timing offset (TO) and carrier frequency offset (CFO), mobility leads to Doppler shifts, and multipath propagation further complicates signal processing. While TO stems from symbol timing misalignment and CFO arises from oscillator mismatch or relative motion, both can severely degrade system performance. Unlike conventional communication systems that treat delay and Doppler simply as interference to be mitigated, ISAC systems must precisely estimate these parameters to extract sensing information. Hence, ISAC introduces a unique paradigm shift in handling these distortions. However, achieving high sensing accuracy is particularly challenging in asynchronous deployments, where the lack of common timing and frequency references necessitates sophisticated estimation strategies.

To cope with such impairments, a variety of estimation and compensation methods have been developed. In classical communication systems, CFO is often effectively mitigated by calculating the auto-correlation of training sequences \cite{schmidl1996low}. Alternatively, from a Bayesian learning perspective, joint channel and CFO estimation can be cast as a maximum-likelihood problem and addressed effectively using the space-alternating generalized expectation-maximization (SAGE) method \cite{pun2007iterative, pham2008joint}. To further improve accuracy, prior statistical knowledge of the unknown parameters can be exploited through belief propagation for maximum a posteriori estimation \cite{sun2021massive, wei2022accurate}. In this context, \cite{ito2024joint} proposes a joint channel and CFO estimation method for multiuser MIMO-OFDM systems based on a Bernoulli--Gaussian prior for the channel and a Gaussian mixture model for CFO. Beyond statistical priors, the inherent structure of the signal can be exploited; specifically, since the phase shift induced by CFO exhibits sparsity in the angle-delay domain, bilinear message-passing algorithms have been proposed to solve the resulting sparsity recovery problem \cite{myers2019message}. 

For asynchronous ISAC systems, several approaches have been developed to address clock mismatch. One straightforward solution is to employ a GPS-disciplined oscillator (GPSDO) for timing synchronization in distributed radar systems \cite{zhang2022integration}. However, this approach relies on costly hardware and may not be practical for low-cost deployments. Alternatively, some methods exploit the observation that TO and CFO are common across antennas. For example, cross-antenna cross-correlation (CACC) has been used to mitigate the effects of TO and CFO \cite{ni2021uplink, li2022csi}. Similarly, \cite{ni2023uplink, zeng2019farsense, zeng2020multisense} computes the ratios of channel state information between a reference antenna and the remaining antennas, referred to as the cross-antenna signal ratio (CASR), to compensate for clock asynchronization. More recently, \cite{11071294} extended this line of research to bistatic hybrid-array uplink sensing by first constructing a clean line-of-sight reference signal, then applying a signal-ratio operation to cancel CFO and TO, and finally performing low-complexity, high-resolution joint Doppler-delay estimation using the AoA-Based 2D-FFT-MUSIC (AB2FM) algorithm. Another line of work explicitly estimates TO and CFO prior to sensing and then compensates for their effects. In contrast to CACC and CASR, this approach neither depends on antenna-specific received signals nor assumes the presence of a dominant line-of-sight path between the transmitter and receiver. For instance, \cite{jump} estimates TO through correlation and estimates CFO under the assumption that a reference static path exists. Likewise, in \cite{zhao2023multiple, ventura2026asymov}, the angle of arrival (AoA) is first estimated and then used to distinguish static and dynamic paths for clock asynchronization compensation and Doppler estimation.

In addition to synchronization mismatch, ISAC systems must also contend with the limited pilot resources available for reliable sensing and parameter estimation. Most existing ISAC studies rely primarily on pilot-centric designs, which restrict sensing to known reference signals and consequently underutilize the rich information embedded in data-carrying symbols, even though pilots typically occupy only a small fraction of the transmission frame \cite{li2026rethinking, xu2025exploiting}. Moreover, recent ISAC studies suggest that pilot sequences alone may be insufficient for robust high-resolution sensing \cite{11071294, brunner2024bistatic}. These observations motivate the use of detected data symbols for sensing, could be referred to as data-aided sensing, as a promising means of enhancing estimation performance without sacrificing communication throughput or incurring additional signaling overhead \cite{park2017expectation, nassirpour2025variational_}. However, the blind reuse of data symbols can degrade system performance if symbol detection errors propagate into the sensing algorithms \cite{brunner2024bistatic}. Recent work has begun addressing this trade-off; for instance, \cite{xu2025exploiting} demonstrates that reusing detected data can significantly enhance sensing accuracy while maintaining precoding gains. Furthermore, specialized architectures are being developed to manage this complexity, such as the two-stage deep learning framework in \cite{hu2024isac}. This approach first employs a neural network for data detection and subsequently utilizes those detected symbols to estimate the Angle of Arrival (AoA) and time delay via the Multiple Signal Classification (MUSIC) algorithm. To solve the transmitter-parameter association problem in orthogonal time frequency space vehicular networks, \cite{yang2024sensing} proposes a constrained bilinear recovery model. Their message passing approach enables high-accuracy joint channel estimation, signal detection, and sensing association.

Although exploiting data symbols can improve efficiency, it introduces significant computational complexity and high-dimensional estimation challenges that necessitate advanced statistical frameworks. In this context, sparse Bayesian learning (SBL) has recently emerged as a powerful paradigm for sensing-oriented parameter estimation in mmWave and OFDM-ISAC systems. In particular, SBL has been adopted for clutter-aware joint angle and velocity estimation \cite{sbl}, and has also been applied to extended-target localization in the presence of Doppler-induced inter-carrier interference (ICI) and clutter \cite{sbl_2,sbl_3}. Despite these advancements, conventional SBL-based methods typically rely on high-resolution dictionaries, which can impose a significant storage burden and computational overhead in large-scale systems. Furthermore, existing studies remain primarily localization-oriented and are largely developed for radar or monostatic settings, rather than a fully asynchronous uplink OFDM-ISAC receiver. As a result, the coupled effects of moving targets, static scatterers, TO, CFO, and unknown communication data have not yet been jointly addressed within a unified probabilistic framework. To bridge this gap, we propose an efficient mean-field variational Bayesian (VB) framework for bistatic uplink OFDM-ISAC systems, where all synchronization, propagation, and communication-related unknowns are modeled as random variables and inferred jointly within a single receiver architecture.  

Our contributions can be summarized as follows:

\begin{itemize}
    \item We propose a mean-field VB framework for asynchronous bistatic uplink OFDM-ISAC, in which TO, CFO, propagation parameters, and transmitted data symbols are jointly inferred in a unified probabilistic receiver, thereby enabling integrated sensing, channel estimation, and data-aided detection under limited reference-signal resources.
    
    \item In the proposed framework, all unknown parameters, including complex path gains, angles of arrival (AoAs), delays, Doppler shifts, TO, and CFO, are modeled as random variables with suitable prior distributions. In particular, von Mises distributions are employed for the angular parameters and the phase terms associated with delay/TO and Doppler/CFO, while Gaussian distributions are adopted for the complex path gains. This modeling choice leads to closed-form updates for all latent variables.

    \item We assess the proposed method using symbol error rate (SER) and channel-parameter normalized mean squared error (NMSE). Simulation results show that the proposed method consistently outperforms the SBL, AB2FM, and SAGE baselines in all evaluated metrics. In addition, the results confirm that leveraging data payload symbols can significantly enhance estimation accuracy without sacrificing communication performance.
\end{itemize}

The remainder of this paper is organized as follows. Section II presents the system model and formulates the problem. Section III discusses the unambiguous delay and Doppler ranges. Section IV introduces the variational Bayesian framework and details the proposed solution. Section V presents the numerical results and evaluates the effectiveness of the proposed method. Finally, Section VI concludes the paper.

\section{System Model and Problem Formulation}

This section presents the signal model and problem formulation for the considered asynchronous uplink OFDM-ISAC system. We first describe the bistatic uplink transmission model, where the BS receives both the direct propagation path and multiple reflected paths from static scatterers and moving targets. Then, we introduce the UPA response model and characterize TO and CFO. Finally, we formulate the objective of joint data detection, channel estimation, synchronization, and sensing-parameter estimation.

\subsection{System Model}
\begin{figure}[t]
  \centering
  \includegraphics[width=1.0\linewidth]{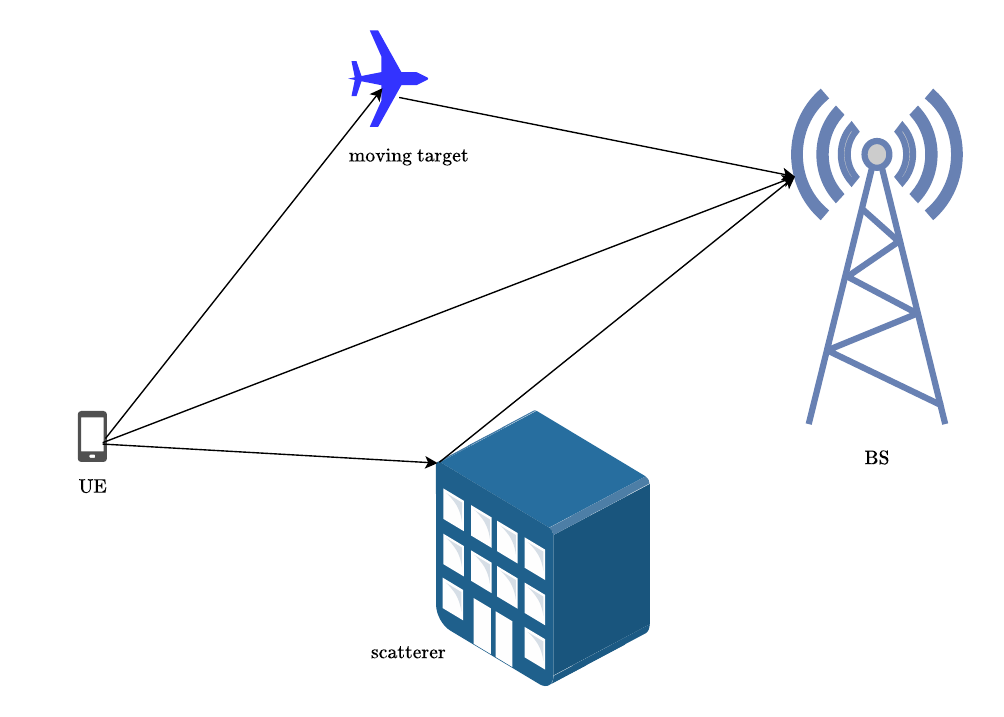} 
  \caption{will be replaced.}
  \label{fig:bistatic_photo}
\end{figure}

We consider a bistatic uplink OFDM-ISAC system, as illustrated in 
Fig.~\ref{fig:bistatic_photo}. The UE transmits an OFDM frame containing both pilot and data symbols, while the BS receives a superposition of the direct path and multiple reflected propagation components. The BS is equipped with a UPA consisting of $N_H$ and $N_V$ antenna elements along the horizontal and vertical directions, respectively, with a total number of receive antennas 
$M=N_HN_V$. The propagation environment contains one moving target and several static scatterers between the UE and the BS. 

Each coherent processing interval (CPI) contains \(K\) OFDM symbols over \(N\) orthogonal subcarriers. Let \((n,k)\) denote the resource element (RE) corresponding to the \(n\)-th subcarrier and the \(k\)-th OFDM symbol, where
\[
n\in\{0,1,\ldots,N-1\}, \qquad
k\in\{0,1,\ldots,K-1\}.
\]
Let \(\Delta_f\) denote the subcarrier spacing, while the duration of each OFDM symbol is
\[
T_{\mathrm{sym}}=\frac{1}{\Delta_f}+T_{\mathrm{CP}},
\]
where \(T_{\mathrm{CP}}\) is the cyclic-prefix duration. The transmitted symbol on RE \((n,k)\) is given by
\begin{equation}
x_{n,k}=
\begin{cases}
x_{n,k}^{(p)}, & (n,k)\in\mathcal{P},\\
x_{n,k}^{(d)}\in\mathcal{X}, & (n,k)\in\mathcal{D},
\end{cases}
\label{eq:xnk}
\end{equation}
where \(\mathcal{P}\) and \(\mathcal{D}\) denote the pilot and data sets, respectively, and \(\mathcal{X}\) is the data constellation.

After cyclic-prefix removal and FFT, the received array snapshot at the BS on resource element (RE) \((n,k)\) can be expressed as
\begin{equation}
\begin{split}
\mathbf{y}_{n,k}
&=
x_{n,k}\,\xi_{n,k}
\sum_{\ell=1}^{L}
\alpha_\ell
\e^{-\ji2\pi n\Delta_f\tau_\ell + \ji2\pi \nu_\ell kT_{\mathrm{sym}}}
\mathbf{b}(\omega_\ell,\psi_\ell)
\\
&\quad
+\mathbf{w}_{n,k},
\end{split}
\label{eq:rx_model}
\end{equation}
where \(\mathbf{y}_{n,k}\in\mathbb{C}^{M}\) denotes the received signal vector and \(\mathbf{w}_{n,k}\sim\mathcal{CN}(\mathbf{0},N_0^{-1}\mathbf{I}_M)\) is the additive white Gaussian noise vector. In addition, \(\alpha_\ell\in\mathbb{C}\), \(\tau_\ell\), and \(\nu_\ell\) represent the complex gain, delay, and Doppler shift of the \(\ell\)-th resolvable path, respectively, while \(\mathbf{b}(\omega_\ell,\psi_\ell)\in\mathbb{C}^{M}\) denotes the UPA steering vector associated with that path. Let the \(i\)-th antenna element be indexed by
\[
m_i=\mathrm{mod}(i-1,N_H), \qquad
n_i=\left\lfloor\frac{i-1}{N_H}\right\rfloor,
\]
so that its position is given by
\[
\mathbf{p}_i=
\begin{bmatrix}
0 & m_i\Delta & n_i\Delta
\end{bmatrix}^{\mathsf T}.
\]
for a plane wave impinging from direction \((\theta_\ell,\varphi_\ell)\), the corresponding array response at the \(i\)-th antenna element is
\[
b_i(\omega_\ell,\psi_\ell)
=
\exp\!\left(\ji\frac{2\pi}{\lambda}\Delta\bigl(m_i\cos\theta_\ell\sin\varphi_\ell+n_i\sin\theta_\ell\bigr)\right),
\]
by defining
\[
\omega_\ell \triangleq \frac{2\pi}{\lambda}\Delta\cos\theta_\ell\sin\varphi_\ell,
\qquad
\psi_\ell \triangleq \frac{2\pi}{\lambda}\Delta\sin\theta_\ell,
\]
the array response can be written as
\[
b_i(\omega_\ell,\psi_\ell)=\e^{\ji\omega_\ell m_i}\e^{\ji\psi_\ell n_i},
\]
accordingly, with
\[
a_i(\omega_\ell)\triangleq \e^{\ji\omega_\ell m_i}, \qquad
c_i(\psi_\ell)\triangleq \e^{\ji\psi_\ell n_i},
\]
the steering vector admits the separable form
\[
\mathbf{b}(\omega_\ell,\psi_\ell)=\mathbf{a}(\omega_\ell)\odot\mathbf{c}(\psi_\ell).
\]

The multiplicative factor \(\xi_{n,k}\) models the residual synchronization mismatch and is expressed as
\begin{equation}
\xi_{n,k}
=
\e^{-\ji2\pi n\Delta_f\tau_{0,k}}
\e^{\ji2\pi f_{0,k}kT_{\mathrm{sym}}},
\label{eq:xi_nk}
\end{equation}
where \(\tau_{0,k}\) and \(f_{0,k}\) denote the TO and CFO, respectively. The above signal model is established under the following assumptions:
\begin{itemize}
    \item \emph{Common oscillator within each array:} All transmit antennas are driven by a common local oscillator (LO), and all receive antennas share another common LO.
    \item \emph{Quasi-static channel:} The physical path parameters are assumed to remain approximately constant over a single CPI.
    \item \emph{Dominant LoS-path:} In the considered uplink scenario, we assume the presence of a dominant LoS path. In addition, this path is assumed to be among the earliest effective-delay paths. In the sequel, this dominant early-arriving path is used for obtaining TO.
\end{itemize}

The objective of this work is to design a synchronization-aware ISAC receiver that jointly detects the transmitted data symbols $\{x_{n,k}\}$, estimates the multipath channel parameters 
$\{\alpha_\ell,\tau_\ell,\nu_\ell,\omega_\ell,\psi_\ell\}_{\ell=1}^{L}$, and 
recovers the residual synchronization parameters $\tau_0$ and $f_0$. These 
estimates are then used for both reliable communication and accurate sensing, including target localization and velocity estimation.

\section{Unambiguous Delay and Doppler Ranges}

The delay and Doppler characteristics of the $\ell$-th propagation path are represented through the discrete phase sequences
\begin{equation}
    g_\ell[n] = \e^{-\ji2\pi n\Delta f\tau_\ell},
\end{equation}
and
\begin{equation}
    u_\ell[k] = \e^{\ji2\pi k\nu_\ell T_{\rm sym}},
\end{equation}
respectively, where $\tau_\ell$ and $\nu_\ell$ denote the path delay and Doppler shift. The identifiability of these parameters is fundamentally determined by the periodicity of the associated complex exponentials. As a result, delay and Doppler can only be uniquely identified within their corresponding unambiguous intervals.

\subsection{Delay Ambiguity} 
We first consider the delay-dependent term $g_\ell[n]$. Two distinct delays, $\tau_\ell$ and $\tau_\ell'$, are indistinguishable if they generate the same phase progression over all subcarriers, namely,
\begin{equation}
    \e^{-\ji2\pi n\Delta f\tau_\ell} = \e^{-\ji2\pi n\Delta f\tau_\ell'}, \quad \forall n.
\end{equation}
This condition is equivalent to
\begin{equation}
    \e^{-\ji2\pi n\Delta f(\tau_\ell-\tau_\ell')} = 1, \quad \forall n,
\end{equation}
which requires
\begin{equation}
    \Delta f(\tau_\ell-\tau_\ell') \in \mathbb{Z}.
\end{equation}
Therefore, the delay parameter is periodic with period
\begin{equation}
    \tau_{\text{per}} = \frac{1}{\Delta f}.
\end{equation}
To ensure uniqueness, $\tau_\ell$ must be restricted to an interval of width $1/\Delta f$. Since physical propagation delays are non-negative, the unambiguous delay region is naturally chosen as
\begin{equation}
    \tau_\ell \in \left[0, \frac{1}{\Delta f}\right).
\end{equation}

\subsection{Doppler Ambiguity}
A similar argument applies to the Doppler-dependent term $q_\ell[k]$. Two Doppler shifts, $\nu_\ell$ and $\nu_\ell'$, are indistinguishable if
\begin{equation}
    \e^{\ji2\pi k\nu_\ell T_{\rm sym}} = \e^{\ji2\pi k\nu_\ell' T_{\rm sym}}, \quad \forall k.
\end{equation}
This implies
\begin{equation}
    (\nu_\ell-\nu_\ell')T_{\rm sym} \in \mathbb{Z},
\end{equation}
from which the Doppler periodicity follows as
\begin{equation}
    \nu_{\text{per}} = \frac{1}{T_{\rm sym}}.
\end{equation}
Accordingly, the Doppler shift can only be uniquely identified within an interval of width $1/T_{\rm sym}$. A common symmetric choice for the unambiguous Doppler range is
\begin{equation}
    \nu_\ell \in \left[-\frac{1}{2T_{\rm sym}}, \frac{1}{2T_{\rm sym}}\right).
\end{equation}

\subsection{Impact of Pilot Subsampling}
The above ambiguity limits are further affected during the pilot-based initialization stage, where observations are available only on every $D_f$-th pilot subcarrier. Letting $n = pD_f$, with $p$ denoting the pilot index, the delay-dependent phase term becomes
\begin{equation}
    g_\ell[p] = \e^{-\ji2\pi pD_f\Delta f\tau_\ell}.
\end{equation}
Under this subsampled structure, two delays are indistinguishable if
\begin{equation}
    D_f\Delta f(\tau_\ell-\tau_\ell') \in \mathbb{Z},
\end{equation}
which leads to a reduced delay periodicity given by
\begin{equation}
    \tau_{\text{per}}^{(p)} = \frac{1}{D_f\Delta f}.
\end{equation}
Hence, when only pilot subcarriers are used for initialization, the unambiguous delay interval is narrowed to
\begin{equation}
    \tau_\ell \in \left[0, \frac{1}{D_f\Delta f}\right).
\end{equation}
This shows that pilot subsampling reduces the delay identifiability range, which may lead to ambiguity in the initialization stage when the true delay exceeds this interval.

\section{Variational Bayesian Inference}
In this section, we develop a mean-field VB algorithm for the considered asynchronous uplink OFDM-ISAC system. The proposed inference framework jointly estimates the transmitted data symbols, multipath channel parameters, and residual synchronization parameters, thereby enabling data-aided sensing under TO and CFO impairments. Specifically, we first briefly review the VB framework and introduce the adopted mean-field factorization. Then, based on the proposed probabilistic model, we derive tractable update rules for all latent variables, including the data symbols, path gains, sparsity precisions, spatial frequencies, delay-Doppler phase parameters, synchronization offsets, and noise precision.
\subsection{Background on VB}
This section briefly reviews variational Bayes (VB). Let $\mb{y}$ drepresent the observed variables, while $\mb{x}$ denotes the latent variables and model parameters. Since the posterior $p(\mb{x}\mid\mb{y})$ is generally intractable, VB approximates it by a distribution $q(\mb{x})\in\mathcal Q$ obtained from \cite{bishop2006pattern, wainwright2008graphical}
\begin{eqnarray}
    q^{\star}(\mathbf{x}) 
    &=& \arg\min_{q(\mathbf{x}) \in \mathcal{Q}} \;
    \mathrm{KL}\!\left( q(\mathbf{x}) \,\|\, p(\mathbf{x} \mid \mathbf{y}) \right)
\end{eqnarray}
which is equivalent to maximizing the evidence lower bound (ELBO)
\begin{eqnarray}
		\mr{ELBO}(q) =  \E_{q(\mb{x})} \big[\ln p(\mb{x},\mb{y})\big] - \E_{q(\mb{x})} \big[\ln q(\mb{x}) \big].
	\end{eqnarray}

Under the mean-field assumption
\begin{align}
q(\mb{x})=\prod_{i=1}^m q_i(x_i),
\label{mean-field}
\end{align}
the optimal factor is given by
\begin{align}
q_i^\star(x_i)
\propto
\exp\!\left\{
\left\langle
\ln p(\mb{x},\mb{y})
\right\rangle_{q_{-i}}
\right\},
\end{align}
where $q_{-i}=\prod_{\ji\neq i}q_j(x_j)$. Sequentially updating $\{q_i(x_i)\}_{i=1}^m$ yields the coordinate-ascent variational inference (CAVI) algorithm, which monotonically increases the ELBO and converges to a local optimum \cite{bishop2006pattern, wainwright2008graphical}.

\subsection{Proposed Variational Bayesian Inference Algorithm}
\label{subsec:proposed_vb}

We define $\gamma \triangleq 1/N_0$ as the noise precision and treat it as
an unknown random variable to be inferred within the VB framework. Given the
received observations $\mathbf{Y}\triangleq\{\mathbf{y}_{n,k}\}$, the objective
is to jointly infer the transmitted symbols $\mathbf{X}\triangleq\{x_{n,k}\}$,
the path gains $\boldsymbol{\alpha}\triangleq[\alpha_1,\ldots,\alpha_{\hat L}]^T$,
the path-dependent delay-Doppler parameters, the spatial frequencies
$\{\omega_\ell,\psi_\ell\}_{\ell=1}^{\hat L}$, and the synchronization parameters.
For notational convenience, we use the phase-domain variables
$\theta_0$, $\phi_0$, $\theta_\ell$, and $\phi_\ell$, and define
\begin{align}
g_0[n] &\triangleq e^{j n\theta_0},&
u_0[k] &\triangleq e^{j k\phi_0}, \nonumber\\
g_\ell[n] &\triangleq e^{j n\theta_\ell},&
u_\ell[k] &\triangleq e^{j k\phi_\ell}.
\label{eq:phase_factors}
\end{align}
The corresponding variational means are denoted by
\begin{align}
\hat{g}_0[n] &\triangleq \left\langle e^{j n\theta_0}\right\rangle,&
\hat{u}_0[k] &\triangleq \left\langle e^{j k\phi_0}\right\rangle, \nonumber\\
\hat{g}_\ell[n] &\triangleq \left\langle e^{j n\theta_\ell}\right\rangle,&
\hat{u}_\ell[k] &\triangleq \left\langle e^{j k\phi_\ell}\right\rangle.
\label{eq:phase_moments}
\end{align}

Let $\mathbf{b}_\ell\triangleq\mathbf{b}(\omega_\ell,\psi_\ell)$ denote the
array response vector associated with the $\ell$-th path. Then, the channel at
time-frequency index $(n,k)$ is written as
\begin{align}
\mathbf{h}_{n,k}
=
g_0[n]u_0[k]
\sum_{\ell=1}^{\hat L}
\alpha_\ell\mathbf{b}_\ell g_\ell[n]u_\ell[k].
\label{eq:hnk_model}
\end{align}
The received signal is modeled as
\begin{align}
\mathbf{y}_{n,k}
=
x_{n,k}\mathbf{h}_{n,k}+\mathbf{w}_{n,k},
\qquad
\mathbf{w}_{n,k}\sim\mathcal{CN}(\mathbf{0},\gamma^{-1}\mathbf{I}_M).
\label{eq:obs_model}
\end{align}

Following the mean-field VB framework, the posterior distribution is
approximated as
\begin{align}
&p(\mathbf{X},\boldsymbol{\alpha},\boldsymbol{\beta},
\boldsymbol{\theta},\boldsymbol{\phi},
\boldsymbol{\omega},\boldsymbol{\psi},\theta_0,\phi_0,\gamma
\mid \mathbf{Y})
\nonumber\\
&\quad\approx
q(\mathbf{X})q(\gamma)q(\theta_0)q(\phi_0)
\prod_{\ell=1}^{\hat L}
q(\alpha_\ell)q(\beta_\ell)q(\theta_\ell)q(\phi_\ell)
q(\omega_\ell)q(\psi_\ell).
\label{eq:mean_field}
\end{align}
The corresponding joint distribution can be factorized as
\begin{align}
&p(\mathbf{Y},\mathbf{X},\boldsymbol{\alpha},\boldsymbol{\beta},
\boldsymbol{\theta},\boldsymbol{\phi},
\boldsymbol{\omega},\boldsymbol{\psi},\theta_0,\phi_0,\gamma)
\nonumber\\
&\quad =
p(\mathbf{Y}\mid \mathbf{X},\boldsymbol{\alpha},
\boldsymbol{\theta},\boldsymbol{\phi},
\boldsymbol{\omega},\boldsymbol{\psi},\theta_0,\phi_0,\gamma)
p(\mathbf{X})p(\gamma)p(\theta_0)p(\phi_0)
\nonumber\\
&\qquad \times
\prod_{\ell=1}^{\hat L}
p(\alpha_\ell\mid\beta_\ell)p(\beta_\ell)
p(\theta_\ell)p(\phi_\ell)p(\omega_\ell)p(\psi_\ell),
\label{eq:joint_factorization}
\end{align}
where the sparse Gamma-Gaussian prior is imposed as
\begin{align}
p(\alpha_\ell\mid\beta_\ell)
=
\mathcal{CN}(\alpha_\ell;0,\beta_\ell^{-1}),
\qquad
p(\beta_\ell)
=
\mathrm{Gamma}(a_\beta,b_\beta).
\label{eq:gamma_gaussian_prior}
\end{align}

To isolate the contribution of the $\ell$-th path, we define the residual
signal
\begin{align}
\mathbf{r}_{n,k}^{(\ell)}
&\triangleq
\mathbf{y}_{n,k}-x_{n,k}\mathbf{h}_{n,k}^{(-\ell)}
\nonumber\\
&=
x_{n,k}\alpha_\ell\mathbf{b}_\ell
g_0[n]u_0[k]g_\ell[n]u_\ell[k]
+\mathbf{w}_{n,k},
\label{eq:residual_l}
\end{align}
where
\begin{align}
\mathbf{h}_{n,k}^{(-\ell)}
\triangleq
g_0[n]u_0[k]
\sum_{i\neq \ell}
\alpha_i\mathbf{b}_i g_i[n]u_i[k].
\label{eq:h_minus_l}
\end{align}
By construction, $\mathbf{r}_{n,k}^{(\ell)}$ contains only the signal component
associated with the $\ell$-th path plus noise, and is used to decouple the
per-path variational updates.

\emph{1) Updating $q(x_{n,k})$:}
By taking the expectation over the joint distribution of all latent variables other than $x_{n,k}$, we obtain
\begin{equation}
\begin{aligned}
q(x_{n,k})
&\propto
\exp\Biggl\{
\Bigl\langle
\ln p(\mathbf{y}_{n,k}\mid x_{n,k},\mathbf{h}_{n,k},\gamma)
+\ln p(x_{n,k})
\Bigr\rangle_{-x_{n,k}}
\Biggr\}.
\end{aligned}
\label{eq:qx_general}
\end{equation}
For a discrete constellation $\mathcal{X}$, the probability of
$x_{n,k}=s$, $s\in\mathcal{X}$, is given by
\begin{align}
q(x_{n,k}=s)
&\propto
p_S(s)
\exp\left(
2\hat{\gamma}\Re\left\{
s^*\hat{\mathbf{h}}_{n,k}^{H}\mathbf{y}_{n,k}
\right\}
-\hat{\gamma}|s|^2E_{n,k}
\right),
\label{eq:x_update}
\end{align}
where
\begin{align}
\hat{\mathbf{h}}_{n,k}
&\triangleq
\left\langle \mathbf{h}_{n,k}\right\rangle
\nonumber\\
&=
\hat{g}_0[n]\hat{u}_0[k]
\sum_{\ell=1}^{\hat L}
\hat{\alpha}_\ell\hat{\mathbf{b}}_\ell
\hat{g}_\ell[n]\hat{u}_\ell[k],
\label{eq:h_mean}
\end{align}
and
\begin{align}
E_{n,k}
\triangleq
\left\langle \|\mathbf{h}_{n,k}\|^2 \right\rangle .
\label{eq:Enk_def}
\end{align}
Since $|g_0[n]|=|u_0[k]|=1$, the second moment can be expressed as
\begin{align}
E_{n,k}
&=
\sum_{\ell=1}^{\hat L}
\left(|\hat{\alpha}_\ell|^2+\tau_{\alpha_\ell}\right)
\|\hat{\mathbf{b}}_\ell\|^2
\nonumber\\
&\quad+
\sum_{\ell=1}^{\hat L}
\sum_{\substack{i=1\\ i\neq \ell}}^{\hat L}
\hat{\alpha}_\ell\hat{\alpha}_i^*
\hat{\mathbf{b}}_i^H\hat{\mathbf{b}}_\ell
\hat{g}_\ell[n]\hat{g}_i^*[n]
\hat{u}_\ell[k]\hat{u}_i^*[k],
\label{eq:E_nk}
\end{align}
with
\begin{align}
\hat{\alpha}_\ell \triangleq \langle \alpha_\ell\rangle,
\qquad
\tau_{\alpha_\ell}
\triangleq
\left\langle |\alpha_\ell|^2\right\rangle
-
|\hat{\alpha}_\ell|^2.
\label{eq:alpha_stats}
\end{align}

\emph{2) Updating $q(\alpha_\ell)$:}
Computing the expectation of the joint distribution over all latent variables except $\alpha_\ell$, the variational distribution of $\alpha_\ell$
is given by
\begin{equation}
\begin{aligned}
q(\alpha_\ell)
&\propto
\exp\Biggl\{
\Bigl\langle
\ln p\!\left(
\{\mathbf{r}_{n,k}^{(\ell)}\}
\mid
\mathbf{X},\boldsymbol{\alpha},
\boldsymbol{\theta},\boldsymbol{\phi},
\boldsymbol{\omega},\boldsymbol{\psi},
\theta_0,\phi_0,\gamma
\right)
\\
&\quad
+\ln p(\alpha_\ell\mid\beta_\ell)
\Bigr\rangle_{-\alpha_\ell}
\Biggr\}.
\end{aligned}
\label{eq:qalpha_general}
\end{equation}
This yields the complex Gaussian distribution
\begin{align}
q(\alpha_\ell)
=
\mathcal{CN}(\alpha_\ell;\hat{\alpha}_\ell,\tau_{\alpha_\ell}),
\label{eq:qalpha_distribution}
\end{align}
where
\begin{align}
\tau_{\alpha_\ell}
&=
\left(\hat{\gamma}S_\ell+\hat{\beta}_\ell\right)^{-1},
\nonumber\\
\hat{\alpha}_\ell
&=
\tau_{\alpha_\ell}\hat{\gamma}t_\ell.
\label{eq:alpha_update}
\end{align}
Here,
\begin{align}
t_\ell
&\triangleq
\sum_{n,k}
\hat{x}_{n,k}^{*}
\hat{g}_0^*[n]\hat{u}_0^*[k]
\hat{g}_\ell^*[n]\hat{u}_\ell^*[k]
\hat{\mathbf{b}}_\ell^H
\left\langle \mathbf{r}_{n,k}^{(\ell)}\right\rangle,
\nonumber\\
S_\ell
&\triangleq
\sum_{n,k}
v_{n,k}\|\hat{\mathbf{b}}_\ell\|^2
=
\|\hat{\mathbf{b}}_\ell\|^2
\sum_{n,k}v_{n,k},
\label{eq:t_l_S_l}
\end{align}
with
\begin{align}
\hat{x}_{n,k}\triangleq \langle x_{n,k}\rangle,
\qquad
v_{n,k}\triangleq \left\langle |x_{n,k}|^2\right\rangle,
\qquad
\hat{\beta}_\ell\triangleq \langle \beta_\ell\rangle .
\label{eq:x_beta_stats}
\end{align}

\emph{3) Updating $q(\beta_\ell)$:}
Taking the expectation with respect to all latent variables except
$\beta_\ell$, we have
\begin{equation}
\begin{aligned}
q(\beta_\ell)
&\propto
\exp\Biggl\{
\Bigl\langle
\ln p(\alpha_\ell\mid\beta_\ell)
+\ln p(\beta_\ell)
\Bigr\rangle_{-\beta_\ell}
\Biggr\}
\\
&=
\mathrm{Gamma}\left(
a_\beta+1,\;
b_\beta+|\hat{\alpha}_\ell|^2+\tau_{\alpha_\ell}
\right).
\end{aligned}
\label{eq:qbeta_distribution}
\end{equation}
Therefore, the variational mean of $\beta_\ell$ is
\begin{align}
\hat{\beta}_\ell
=
\frac{a_\beta+1}
{b_\beta+|\hat{\alpha}_\ell|^2+\tau_{\alpha_\ell}}.
\label{eq:beta_update}
\end{align}

\emph{4) Updating $q(\gamma)$:}
By taking the expectation over the joint distribution of all latent variables other than $\gamma$,
the variational distribution of $\gamma$ is given by
\begin{equation}
\begin{aligned}
q(\gamma)
&\propto
\exp\Biggl\{
\Bigl\langle
\ln p\!\left(
\mathbf{Y}
\mid
\mathbf{X},\boldsymbol{\alpha},
\boldsymbol{\theta},\boldsymbol{\phi},
\boldsymbol{\omega},\boldsymbol{\psi},
\theta_0,\phi_0,\gamma
\right)
\\
&\quad
+\ln p(\gamma)
\Bigr\rangle_{-\gamma}
\Biggr\}
\\
&=
\mathrm{Gamma}\left(
a_0+M N_{\mathrm{sc}}K_{\mathrm{sym}},\;
b_0+E_{\mathrm{err}}
\right).
\end{aligned}
\label{eq:qgamma_distribution}
\end{equation}
where
\begin{align}
E_{\mathrm{err}}
&\triangleq
\sum_{n,k}
\left\langle
\|\mathbf{y}_{n,k}-x_{n,k}\mathbf{h}_{n,k}\|^2
\right\rangle
\nonumber\\
&=
\sum_{n,k}
\left(
\|\mathbf{y}_{n,k}\|^2
-
2\Re\left\{
\hat{x}_{n,k}^*
\hat{\mathbf{h}}_{n,k}^{H}\mathbf{y}_{n,k}
\right\}
+
v_{n,k}E_{n,k}
\right).
\label{eq:Eerr}
\end{align}
Hence,
\begin{align}
\hat{\gamma}
=
\frac{a_0+M N_{\mathrm{sc}}K_{\mathrm{sym}}}
{b_0+E_{\mathrm{err}}}.
\label{eq:gamma_update}
\end{align}

\emph{5) Updating $q(\omega_\ell)$:}
By taking the expectation over the joint distribution of all latent variables other than $\omega_\ell$, we obtain
\begin{equation}
\begin{aligned}
q(\omega_\ell)
&\propto
\exp\Biggl\{
\Bigl\langle
\ln p\!\left(
\{\mathbf{r}_{n,k}^{(\ell)}\}
\mid
\mathbf{X},\boldsymbol{\alpha},
\boldsymbol{\theta},\boldsymbol{\phi},
\boldsymbol{\omega},\boldsymbol{\psi},
\theta_0,\phi_0,\gamma
\right)
\\
&\quad
+\ln p(\omega_\ell)
\Bigr\rangle_{-\omega_\ell}
\Biggr\}.
\end{aligned}
\label{eq:qomega_general}
\end{equation}
After discarding terms independent of $\omega_\ell$, we have
\begin{align}
\ln q(\omega_\ell)
&\propto
\Re\left\{
2\hat{\gamma}\hat{\alpha}_\ell^*
\mathbf{b}(\omega_\ell,\hat{\psi}_\ell)^H
\mathbf{z}_\ell
\right\}
+
\ln p(\omega_\ell),
\label{eq:qomega_log}
\end{align}
where
\begin{align}
\mathbf{z}_\ell
\triangleq
\sum_{n,k}
\left\langle \mathbf{r}_{n,k}^{(\ell)}\right\rangle
\hat{x}_{n,k}^{*}
\hat{g}_0^*[n]\hat{u}_0^*[k]
\hat{g}_\ell^*[n]\hat{u}_\ell^*[k].
\label{eq:z_l}
\end{align}
Let $m_i$ and $n_i$ denote the horizontal and vertical antenna indices,
respectively. Then,
\begin{align}
\ln q(\omega_\ell)
&\propto
\Re\left\{
2\hat{\gamma}\hat{\alpha}_\ell^*
\sum_i
z_{\ell,i}
e^{-j\hat{\psi}_\ell n_i}
e^{-j\omega_\ell m_i}
\right\}
+
\ln p(\omega_\ell).
\label{eq:qomega_log_expanded}
\end{align}
Grouping the terms with the same horizontal index $m_i$, define
\begin{align}
\eta_\ell^{(\omega)}[m]
\triangleq
2\hat{\gamma}\hat{\alpha}_\ell^*
\sum_{i:m_i=m}
z_{\ell,i}e^{-j\hat{\psi}_\ell n_i}.
\label{eq:eta_omega}
\end{align}
Therefore,
\begin{align}
q(\omega_\ell)
&\propto
p(\omega_\ell)
\exp\left(
\Re\left\{
\sum_m
\eta_\ell^{(\omega)}[m]e^{-jm\omega_\ell}
\right\}
\right).
\label{eq:q_omega}
\end{align}
We approximate $q(\omega_\ell)$ by a single von Mises (VM) distribution with
mean direction $\hat{\omega}_\ell$ and concentration parameter
$\kappa_{\omega_\ell}$. Specifically, we employ the Heuristic~2 procedure
in~\cite{badiu2017variational}, which maps a distribution $q(\omega_\ell)$
to a VM approximation and returns the corresponding mean direction
$\hat{\omega}_\ell$ and concentration parameter $\kappa_{\omega_\ell}$. Its moments are used to
compute the expected steering vector $\hat{\mathbf{a}}(\omega_\ell)$.

\emph{6) Updating $q(\psi_\ell)$:}
Similarly, fixing $\omega_\ell=\hat{\omega}_\ell$ and taking the expectation
with respect to all latent variables except $\psi_\ell$, we obtain
\begin{equation}
\begin{aligned}
q(\psi_\ell)
&\propto
\exp\Biggl\{
\Bigl\langle
\ln p\!\left(
\{\mathbf{r}_{n,k}^{(\ell)}\}
\mid
\mathbf{X},\boldsymbol{\alpha},
\boldsymbol{\theta},\boldsymbol{\phi},
\boldsymbol{\omega},\boldsymbol{\psi},
\theta_0,\phi_0,\gamma
\right)
\\
&\quad
+\ln p(\psi_\ell)
\Bigr\rangle_{-\psi_\ell}
\Biggr\}.
\end{aligned}
\label{eq:qpsi_general}
\end{equation}
After discarding terms independent of $\psi_\ell$, we have
\begin{align}
\ln q(\psi_\ell)
&\propto
\Re\left\{
2\hat{\gamma}\hat{\alpha}_\ell^*
\mathbf{b}(\hat{\omega}_\ell,\psi_\ell)^H
\mathbf{z}_\ell
\right\}
+
\ln p(\psi_\ell).
\label{eq:qpsi_log}
\end{align}
Equivalently,
\begin{align}
\ln q(\psi_\ell)
&\propto
\Re\left\{
2\hat{\gamma}\hat{\alpha}_\ell^*
\sum_i
z_{\ell,i}
e^{-j\hat{\omega}_\ell m_i}
e^{-j\psi_\ell n_i}
\right\}
+
\ln p(\psi_\ell).
\label{eq:qpsi_log_expanded}
\end{align}
Grouping the terms with the same vertical index $n_i$, define
\begin{align}
\eta_\ell^{(\psi)}[n]
\triangleq
2\hat{\gamma}\hat{\alpha}_\ell^*
\sum_{i:n_i=n}
z_{\ell,i}e^{-j\hat{\omega}_\ell m_i}.
\label{eq:eta_psi}
\end{align}
Thus,
\begin{align}
q(\psi_\ell)
&\propto
p(\psi_\ell)
\exp\left(
\Re\left\{
\sum_n
\eta_\ell^{(\psi)}[n]e^{-jn\psi_\ell}
\right\}
\right).
\label{eq:q_psi}
\end{align}
The distribution $q(\psi_\ell)$ is also approximated by a von Mises
distribution, and its moments are used to compute the expected steering vector
$\hat{\mathbf{c}}(\psi_\ell)$.

\emph{7) Updating $q(\theta_\ell)$:}
By taking the expectation over the joint distribution of all latent variables other than $\theta_\ell$, the variational distribution of $\theta_\ell$ is given by
\begin{equation}
\begin{aligned}
q(\theta_\ell)
&\propto
\exp\Biggl\{
\Bigl\langle
\ln p\!\left(
\{\mathbf{r}_{n,k}^{(\ell)}\}
\mid
\mathbf{X},\boldsymbol{\alpha},
\boldsymbol{\theta},\boldsymbol{\phi},
\boldsymbol{\omega},\boldsymbol{\psi},
\theta_0,\phi_0,\gamma
\right)
\\
&\quad
+\ln p(\theta_\ell)
\Bigr\rangle_{-\theta_\ell}
\Biggr\}.
\end{aligned}
\label{eq:qtheta_l_general}
\end{equation}
After expanding the squared norm and removing the terms independent of
$\theta_\ell$, we obtain
\begin{align}
\ln q(\theta_\ell)
&\propto
\Re\left\{
2\hat{\gamma}
\sum_{n,k}
\hat{x}_{n,k}^{*}
\hat{\alpha}_\ell^{*}
\hat{g}_0^*[n]\hat{u}_0^*[k]\hat{u}_\ell^*[k]
\hat{\mathbf{b}}_\ell^H
\left\langle\mathbf{r}_{n,k}^{(\ell)}\right\rangle
e^{-jn\theta_\ell}
\right\}
\nonumber\\
&\quad+
\ln p(\theta_\ell).
\label{eq:qtheta_l_log}
\end{align}
Therefore,
\begin{align}
q(\theta_\ell)
&\propto
p(\theta_\ell)
\exp\left(
\Re\left\{
\sum_n
\eta_\ell^{(\theta)}[n]e^{-jn\theta_\ell}
\right\}
\right),
\label{eq:q_theta_l}
\end{align}
where
\begin{align}
\eta_\ell^{(\theta)}[n]
\triangleq
2\hat{\gamma}
\sum_k
\hat{x}_{n,k}^{*}
\hat{\alpha}_\ell^{*}
\hat{g}_0^*[n]\hat{u}_0^*[k]\hat{u}_\ell^*[k]
\hat{\mathbf{b}}_\ell^H
\left\langle\mathbf{r}_{n,k}^{(\ell)}\right\rangle .
\label{eq:eta_theta_l}
\end{align}

\emph{8) Updating $q(\phi_\ell)$:}
Taking the expectation with respect to all latent variables except
$\phi_\ell$, the variational distribution of $\phi_\ell$ is given by
\begin{equation}
\begin{aligned}
q(\phi_\ell)
&\propto
\exp\Biggl\{
\Bigl\langle
\ln p\!\left(
\{\mathbf{r}_{n,k}^{(\ell)}\}
\mid
\mathbf{X},\boldsymbol{\alpha},
\boldsymbol{\theta},\boldsymbol{\phi},
\boldsymbol{\omega},\boldsymbol{\psi},
\theta_0,\phi_0,\gamma
\right)
\\
&\quad
+\ln p(\phi_\ell)
\Bigr\rangle_{-\phi_\ell}
\Biggr\}.
\end{aligned}
\label{eq:qphi_l_general}
\end{equation}
After removing the terms independent of $\phi_\ell$, we obtain
\begin{align}
\ln q(\phi_\ell)
&\propto
\Re\left\{
2\hat{\gamma}
\sum_{n,k}
\hat{x}_{n,k}^{*}
\hat{\alpha}_\ell^{*}
\hat{g}_0^*[n]\hat{u}_0^*[k]\hat{g}_\ell^*[n]
\hat{\mathbf{b}}_\ell^H
\left\langle\mathbf{r}_{n,k}^{(\ell)}\right\rangle
e^{-jk\phi_\ell}
\right\}
\nonumber\\
&\quad+
\ln p(\phi_\ell).
\label{eq:qphi_l_log}
\end{align}
Equivalently,
\begin{align}
q(\phi_\ell)
&\propto
p(\phi_\ell)
\exp\left(
\Re\left\{
\sum_k
\eta_\ell^{(\phi)}[k]e^{-jk\phi_\ell}
\right\}
\right),
\label{eq:q_phi_l}
\end{align}
where
\begin{align}
\eta_\ell^{(\phi)}[k]
\triangleq
2\hat{\gamma}
\sum_n
\hat{x}_{n,k}^{*}
\hat{\alpha}_\ell^{*}
\hat{g}_0^*[n]\hat{u}_0^*[k]\hat{g}_\ell^*[n]
\hat{\mathbf{b}}_\ell^H
\left\langle\mathbf{r}_{n,k}^{(\ell)}\right\rangle .
\label{eq:eta_phi_l}
\end{align}

\emph{9) Updating $q(\theta_0)$:}
By taking the expectation over the joint distribution of all latent variables other than $\theta_0$, we obtain
\begin{equation}
\begin{aligned}
q(\theta_0)
&\propto
\exp\Biggl\{
\Bigl\langle
\ln p\!\left(
\mathbf{Y}
\mid
\mathbf{X},\boldsymbol{\alpha},
\boldsymbol{\theta},\boldsymbol{\phi},
\boldsymbol{\omega},\boldsymbol{\psi},
\theta_0,\phi_0,\gamma
\right)
\\
&\quad
+\ln p(\theta_0)
\Bigr\rangle_{-\theta_0}
\Biggr\}.
\end{aligned}
\label{eq:qtheta0_general}
\end{equation}
To derive the update, we rewrite the observation model as
\begin{align}
\mathbf{y}_{n,k}
=
x_{n,k}\tilde{\mathbf{h}}_{n,k}^{(\theta_0)}
e^{jn\theta_0}
+
\mathbf{w}_{n,k},
\label{eq:theta0_obs_rewrite}
\end{align}
where
\begin{align}
\tilde{\mathbf{h}}_{n,k}^{(\theta_0)}
\triangleq
u_0[k]
\sum_{\ell=1}^{\hat L}
\alpha_\ell\mathbf{b}_\ell g_\ell[n]u_\ell[k].
\label{eq:theta0_tilde_h}
\end{align}
Since $|e^{jn\theta_0}|^2=1$, the quadratic term is independent of
$\theta_0$. Hence,
\begin{align}
\ln q(\theta_0)
&\propto
\Re\left\{
2\hat{\gamma}
\sum_{n,k}
\hat{x}_{n,k}^{*}
\left(
\bar{\mathbf{h}}_{n,k}^{(\theta_0)}
\right)^H
\mathbf{y}_{n,k}
e^{-jn\theta_0}
\right\}
+
\ln p(\theta_0),
\label{eq:qtheta0_log}
\end{align}
where
\begin{align}
\bar{\mathbf{h}}_{n,k}^{(\theta_0)}
&\triangleq
\left\langle
\tilde{\mathbf{h}}_{n,k}^{(\theta_0)}
\right\rangle_{-\theta_0}
\nonumber\\
&=
\hat{u}_0[k]
\sum_{\ell=1}^{\hat L}
\hat{\alpha}_\ell
\hat{\mathbf{b}}_\ell
\hat{g}_\ell[n]\hat{u}_\ell[k].
\label{eq:theta0_hbar}
\end{align}
Thus,
\begin{align}
q(\theta_0)
&\propto
p(\theta_0)
\exp\left(
\Re\left\{
\sum_n
\eta_0^{(\theta)}[n]e^{-jn\theta_0}
\right\}
\right),
\label{eq:q_theta0}
\end{align}
where
\begin{align}
\eta_0^{(\theta)}[n]
\triangleq
2\hat{\gamma}
\sum_k
\hat{x}_{n,k}^{*}
\left(
\bar{\mathbf{h}}_{n,k}^{(\theta_0)}
\right)^H
\mathbf{y}_{n,k}.
\label{eq:eta_theta0}
\end{align}

\emph{10) Updating $q(\phi_0)$:}
Finally, taking the expectation with respect to all latent variables except
$\phi_0$, we have
\begin{equation}
\begin{aligned}
q(\phi_0)
&\propto
\exp\Biggl\{
\Bigl\langle
\ln p\!\left(
\mathbf{Y}
\mid
\mathbf{X},\boldsymbol{\alpha},
\boldsymbol{\theta},\boldsymbol{\phi},
\boldsymbol{\omega},\boldsymbol{\psi},
\theta_0,\phi_0,\gamma
\right)
\\
&\quad
+\ln p(\phi_0)
\Bigr\rangle_{-\phi_0}
\Biggr\}.
\end{aligned}
\label{eq:qphi0_general}
\end{equation}
We rewrite the observation model as
\begin{align}
\mathbf{y}_{n,k}
=
x_{n,k}\tilde{\mathbf{h}}_{n,k}^{(\phi_0)}
e^{jk\phi_0}
+
\mathbf{w}_{n,k},
\label{eq:phi0_obs_rewrite}
\end{align}
where
\begin{align}
\tilde{\mathbf{h}}_{n,k}^{(\phi_0)}
\triangleq
g_0[n]
\sum_{\ell=1}^{\hat L}
\alpha_\ell\mathbf{b}_\ell g_\ell[n]u_\ell[k].
\label{eq:phi0_tilde_h}
\end{align}
Since $|e^{jk\phi_0}|^2=1$, the quadratic term is independent of
$\phi_0$. Therefore,
\begin{align}
\ln q(\phi_0)
&\propto
\Re\left\{
2\hat{\gamma}
\sum_{n,k}
\hat{x}_{n,k}^{*}
\left(
\bar{\mathbf{h}}_{n,k}^{(\phi_0)}
\right)^H
\mathbf{y}_{n,k}
e^{-jk\phi_0}
\right\}
+
\ln p(\phi_0),
\label{eq:qphi0_log}
\end{align}
where
\begin{align}
\bar{\mathbf{h}}_{n,k}^{(\phi_0)}
&\triangleq
\left\langle
\tilde{\mathbf{h}}_{n,k}^{(\phi_0)}
\right\rangle_{-\phi_0}
\nonumber\\
&=
\hat{g}_0[n]
\sum_{\ell=1}^{\hat L}
\hat{\alpha}_\ell
\hat{\mathbf{b}}_\ell
\hat{g}_\ell[n]\hat{u}_\ell[k].
\label{eq:phi0_hbar}
\end{align}
Hence,
\begin{align}
q(\phi_0)
&\propto
p(\phi_0)
\exp\left(
\Re\left\{
\sum_k
\eta_0^{(\phi)}[k]e^{-jk\phi_0}
\right\}
\right),
\label{eq:q_phi0}
\end{align}
where
\begin{align}
\eta_0^{(\phi)}[k]
\triangleq
2\hat{\gamma}
\sum_n
\hat{x}_{n,k}^{*}
\left(
\bar{\mathbf{h}}_{n,k}^{(\phi_0)}
\right)^H
\mathbf{y}_{n,k}.
\label{eq:eta_phi0}
\end{align}


\begin{algorithm}[t]
\small
\caption{Proposed Asynchronous JED-VB Algorithm}
\label{alg:async_jed_vb}
\begin{algorithmic}[1]
\Require $\mathbf{Y}$, $\mathcal{P}$, $\mathcal{D}$, $\mathcal{X}$, $\hat L$, $I_{\max}$
\Ensure $\hat{\mathbf X}$, $\hat{\mathbf H}$, $\hat{\gamma}$, $\hat{\theta}_0$, $\hat{\phi}_0$, 
$\{\hat{\alpha}_\ell,\hat{\beta}_\ell,\hat{\omega}_\ell,\hat{\psi}_\ell,
\hat{\theta}_\ell,\hat{\phi}_\ell\}_{\ell=1}^{\hat L}$

\Statex
\Statex \textbf{Pilot-based initialization}
\State Extract pilot observations $\mathbf{Y}_{p}$ and set 
$\mathbf{Y}_{p}^{\rm res}\leftarrow \mathbf{Y}_{p}$.
\For{$\ell=1,\ldots,\hat L$}
    \State Initialize 
    $(\hat{\omega}_\ell,\hat{\psi}_\ell,\hat{\theta}_\ell,\hat{\phi}_\ell,
    \hat{\alpha}_\ell,\hat{\mathbf b}_\ell)$ from $\mathbf{Y}_{p}^{\rm res}$.
    \State Remove the initialized path contribution from $\mathbf{Y}_{p}^{\rm res}$.
\EndFor
\State Refine $\hat{\boldsymbol{\alpha}}$ by ridge LS over pilot REs.
\State Initialize $\hat{\beta}_\ell$, $\hat{\tau}_{\alpha_\ell}$, $\hat{\gamma}$, 
$\hat{x}_{n,k}$, $\hat{\theta}_0$, and $\hat{\phi}_0$.
\State Construct the initial channel estimate $\hat{\mathbf H}$ using \eqref{eq:h_mean}.

\Statex
\Statex \textbf{Joint estimation and detection}
\For{$i=1,\ldots,I_{\max}$}
    \State Reorder paths according to $|\hat{\alpha}_\ell|^2+\hat{\tau}_{\alpha_\ell}$.

    \State Update $q(x_{n,k})$ for all data REs using 
    \eqref{eq:x_update}--\eqref{eq:E_nk}.

    \State Update $q(\theta_0)$ and $q(\phi_0)$ using 
    \eqref{eq:q_theta0} and 
    \eqref{eq:q_phi0}.

    \For{$\ell=1,\ldots,\hat L$}
        \State Form $\langle\mathbf r_{n,k}^{(\ell)}\rangle$ and compute 
        $\mathbf z_\ell$ using \eqref{eq:residual_l} and \eqref{eq:z_l}.

        \State Update $q(\omega_\ell)$ and $q(\psi_\ell)$ using 
        \eqref{eq:q_omega} and 
        \eqref{eq:q_psi}.

        \State Update $q(\theta_\ell)$ and $q(\phi_\ell)$ using 
        \eqref{eq:q_theta_l} and 
        \eqref{eq:q_phi_l}.

        \State Update $q(\alpha_\ell)$ and $q(\beta_\ell)$ using 
        \eqref{eq:alpha_update} and \eqref{eq:beta_update}.
    \EndFor

    \State Reconstruct $\hat{\mathbf H}$ using \eqref{eq:h_mean}.
    \State Update $q(\gamma)$ using \eqref{eq:Eerr}--\eqref{eq:gamma_update}.
\EndFor

\end{algorithmic}
\end{algorithm}

\section{Simulation Results}

\begin{table}[!t]
\caption{Simulation Parameters}
\label{tab:SimPara}
\centering
\begin{tabular}{|p{19em}|p{10em}|}
\hline
BS array size $(N_H,N_V)$ & $(8,8)$ \\ \hline
Total number of BS antennas $M=N_HN_V$ & $64$ \\ \hline
Number of OFDM subcarriers $N_{\rm sc}$ & $1024$ \\ \hline
Number of OFDM symbols $K_{\rm sym}$ & $10$ \\ \hline
System bandwidth $B$ & $100$ MHz \\ \hline
OFDM symbol duration $T_{\rm sym}$ & $1/\Delta f=10.24~\mu$s \\ \hline
Pilot spacing $D_f$ & $16$ \\ \hline
Number of pilot subcarriers $N_p$ & $64$ \\ \hline
Number of true channel paths $L_{\rm true}$ & $6$ \\ \hline
Number of candidate paths $L$ & $12$ \\ \hline
Number of moving targets  & $2$ \\ \hline
Elevation angle $\psi$ & $\mathcal U(-\pi/3,\pi/3)$ \\ \hline
Azimuth angle $\omega$ & $\mathcal U(-\pi/3,\pi/3)$ \\ \hline
Delay $\tau$ & $\mathcal U(0,\tau_{\max})$ \\ \hline
Maximum delay $\tau_{\max}$ & $\dfrac{0.9}{2D_f\Delta f}=288$ ns \\ \hline
Carrier frequency $f_c$ & $15$ GHz \\ \hline
Maximum speed $v_{\max}$ & $50$ m/s \\ \hline
Maximum Doppler $\nu_{\max}$ & $v_{\max}/\lambda_c = 2.5$ kHz \\ \hline
Elevation angle $\theta$ & $\mathcal U(-\pi/3,\pi/3)$ \\ \hline
Azimuth angle $\phi$ & $\mathcal U(-\pi/3,\pi/3)$ \\ \hline
Relative path delay $\tau_t$ & $\mathcal U(0.152~\mu\text{s},\,0.288~\mu\text{s})$ \\ \hline
Doppler $\nu_t$ for static paths & $0$ \\ \hline
Doppler $\nu_t$ for moving path & $\mathcal U(-\nu_{\max},\nu_{\max})$ \\ \hline
\end{tabular}
\end{table}

In this section, simulation results are provided to evaluate the performance of the proposed approach in asynchronous uplink OFDM-ISAC systems. The data-detection performance is measured by the BER, while channel reconstruction is assessed by
\begin{equation}
\mathrm{NMSE}(\widehat{\mathbf h}) =
10\log_{10}\!\left(
\frac{\|\mathbf h-\widehat{\mathbf h}\|_2^2}{\|\mathbf h\|_2^2}
\right),
\end{equation}
where $\mathbf h$ and $\widehat{\mathbf h}$ denote the true and estimated channels, respectively. For the physical parameters, we report
\begin{equation}
\mathrm{MSE}(\widehat{\mathbf z}) =
10\log_{10}\!\left(
\|\mathbf z-\widehat{\mathbf z}\|_2^2
\right),
\end{equation}
where $\mathbf z$ and $\widehat{\mathbf z}$ denote the true and estimated parameter vectors, respectively. This metric is evaluated for the path gains, delays, Doppler shifts, azimuth angles, and elevation angles, denoted by $\widehat{\boldsymbol\alpha}$, $\widehat{\boldsymbol\tau}$, $\widehat{\boldsymbol\nu}$, $\widehat{\boldsymbol\omega}$, and $\widehat{\boldsymbol\psi}$, respectively. The simulation parameters used throughout this study are summarized in Table~\ref{tab:SimPara}. The proposed method is compared with the following baselines:

\begin{itemize}
    \item SBL\cite{sbl}: SBL models the received pilot signal with a sparse parametric dictionary over angle, delay phase, and Doppler phase, and refines the dictionary grid points iteratively in an off-grid Bayesian inference procedure. In our asynchronous adaptation, the algorithm first estimates the effective path parameters, after which TO $\tau_0$ is recovered by LoS anchoring and CFO $\nu_0$ is estimated from the static-path cluster to recover the physical Dopplers.
    
    \item SAGE \cite{pham2008joint}: SAGE is adapted to the asynchronous bistatic setting as a space-alternating generalized EM estimator with sequential conditional updates of the effective path parameters. The common CFO is updated once per iteration from the static-path cluster, while the TO remains absorbed in the effective delays and is recovered afterward via LoS anchoring.

    \item AB2FM \cite{11071294}: AB2FM first estimates the AoAs and constructs a clean LoS reference signal. The common CFO and TO are then suppressed through a signal-ratio operation, after which AB2FM jointly estimates the Doppler and the delay relative to the LoS path using the pre-estimated AoAs.

    \item VB pilot: This is a reduced version of the proposed method that excludes the data-detection stage. It relies only on pilot observations for channel-parameter estimation.
    
\end{itemize}

\begin{figure}[!th]
	\centering
\includegraphics[width=1.0\linewidth]{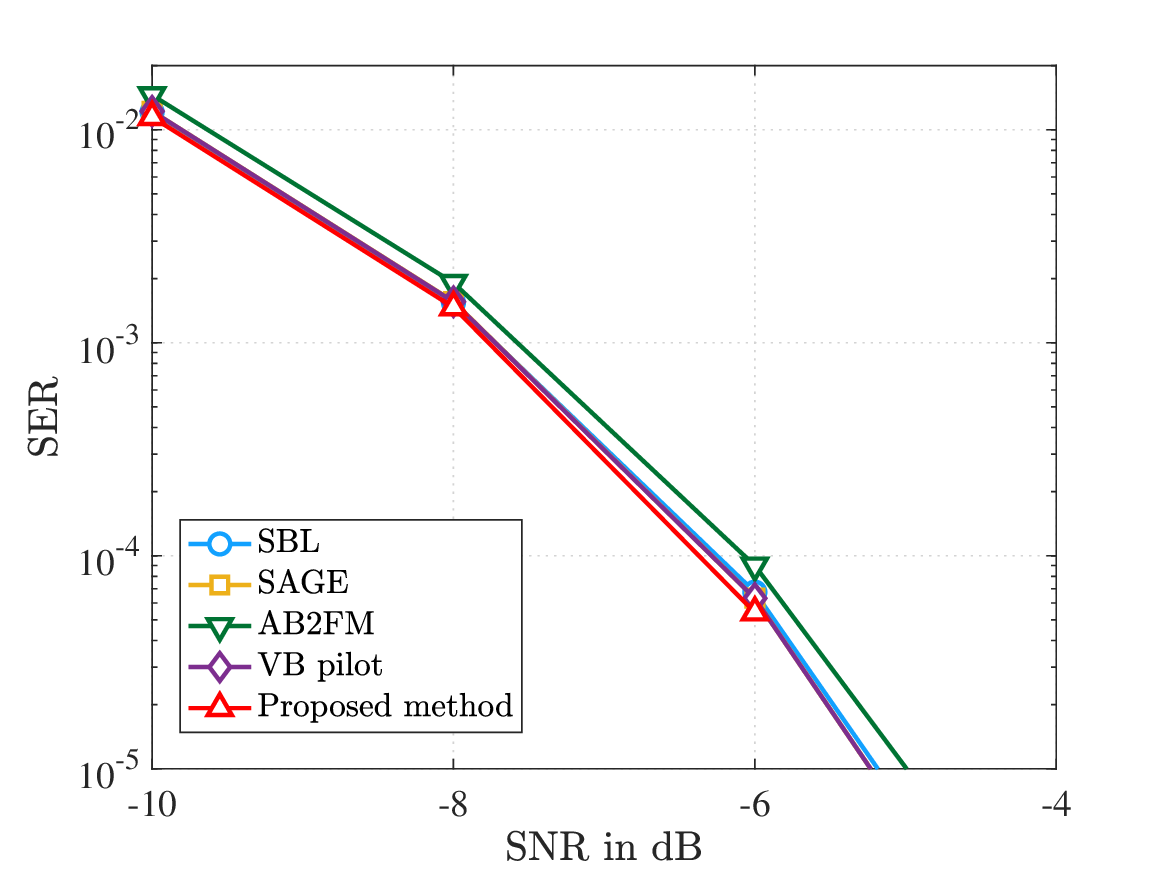}
	\caption{SER performance versus SNR.}
	\label{SER}
\end{figure}

Fig.~\ref{SER} presents the SER performance versus SNR. All considered methods 
achieve reliable data detection in the simulated SNR range. Nevertheless, the 
proposed method consistently attains a slightly lower SER than the baselines, 
mainly due to its channel refinement stage. By jointly refining the channel 
estimate and detecting the transmitted symbols, the proposed framework improves 
the reliability of both channel estimation and data detection. 

\begin{figure}[!th]
	\centering
\includegraphics[width=1.0\linewidth]{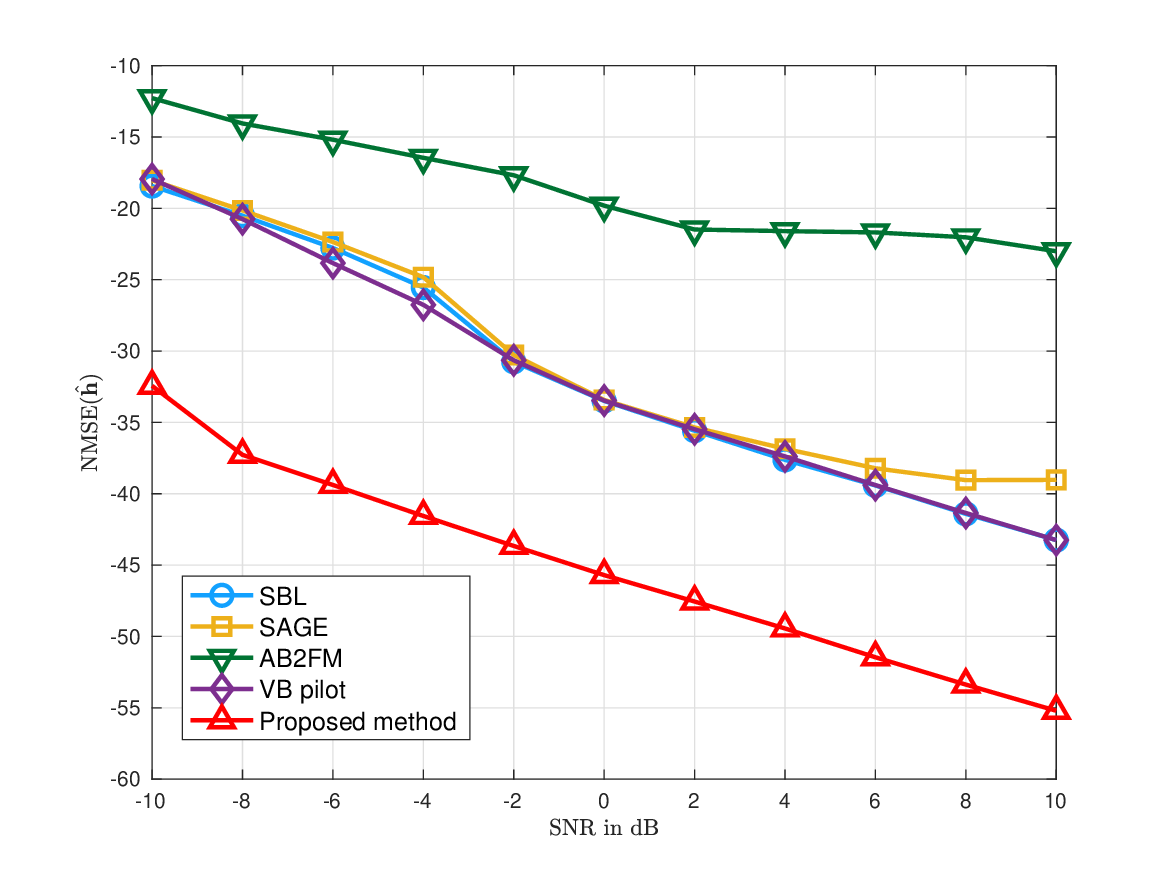}
	\caption{Channel reconstruction NMSE versus SNR for different estimation schemes.}
	\label{channel}
\end{figure}

Fig.~\ref{channel} presents the channel reconstruction NMSE versus SNR. The 
pilot-only VB method already provides promising performance compared with SAGE, 
SBL, and AB2FM, indicating the effectiveness of the proposed VB inference 
framework for channel estimation. After incorporating the detected data symbols 
in the refinement stage, the proposed method achieves a substantial performance 
improvement, with an approximately $10$ dB NMSE gain over the pilot-only case. 
This confirms that reliable data detection can provide additional information for 
channel refinement, thereby enhancing the accuracy of channel estimation.

\begin{figure}[!th]
	\centering
\includegraphics[width=1.0\linewidth]{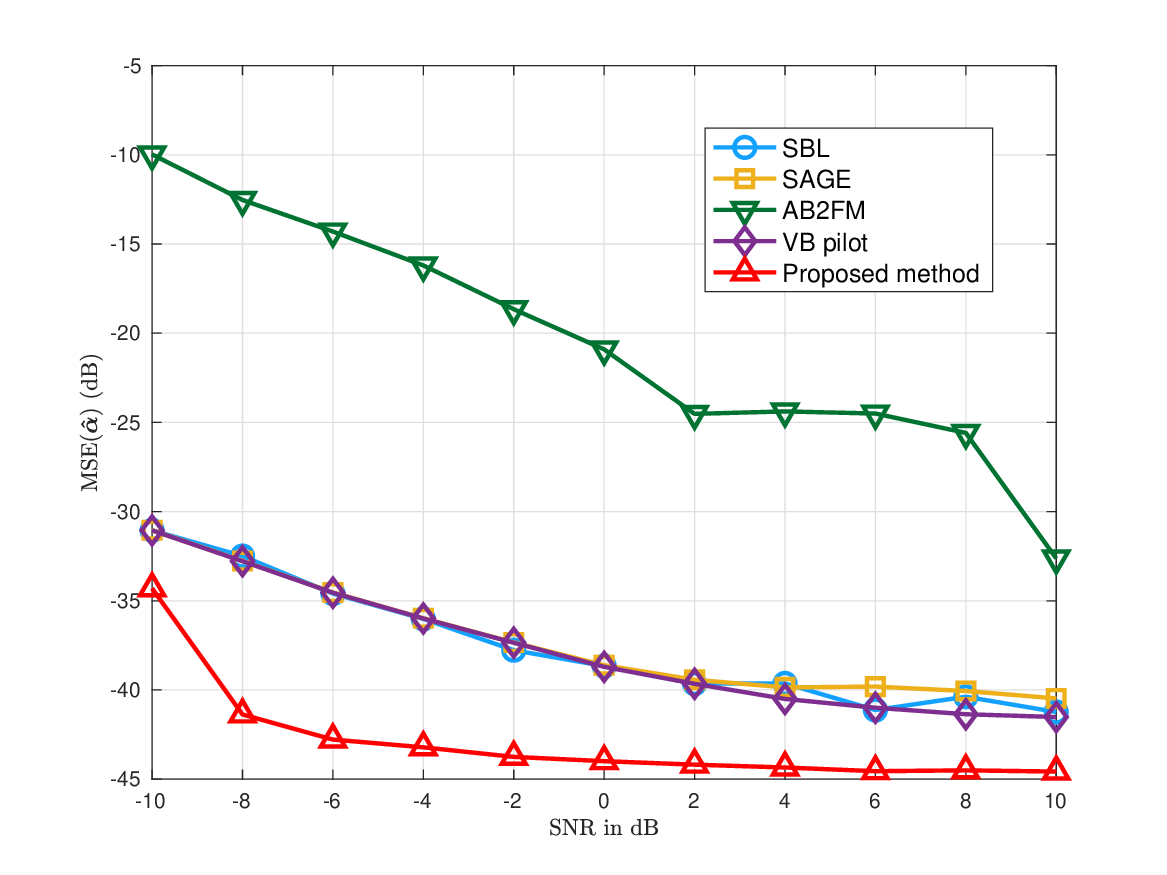}
	\caption{Complex path-gain estimation MSE versus SNR.}
	\label{gain}
\end{figure}

Fig.~\ref{gain} presents the MSE of complex path-gain estimation as a function 
of SNR. The pilot-only VB method achieves performance comparable to SAGE and SBL, 
which is expected since these approaches all rely on statistical inference to 
exploit the sparse multipath structure. By incorporating detected data symbols 
into the refinement stage, the proposed JED-VB method substantially improves the 
path-gain estimation accuracy and consistently outperforms all baselines.

\begin{figure}[!th]
	\centering
\includegraphics[width=1.0\linewidth]{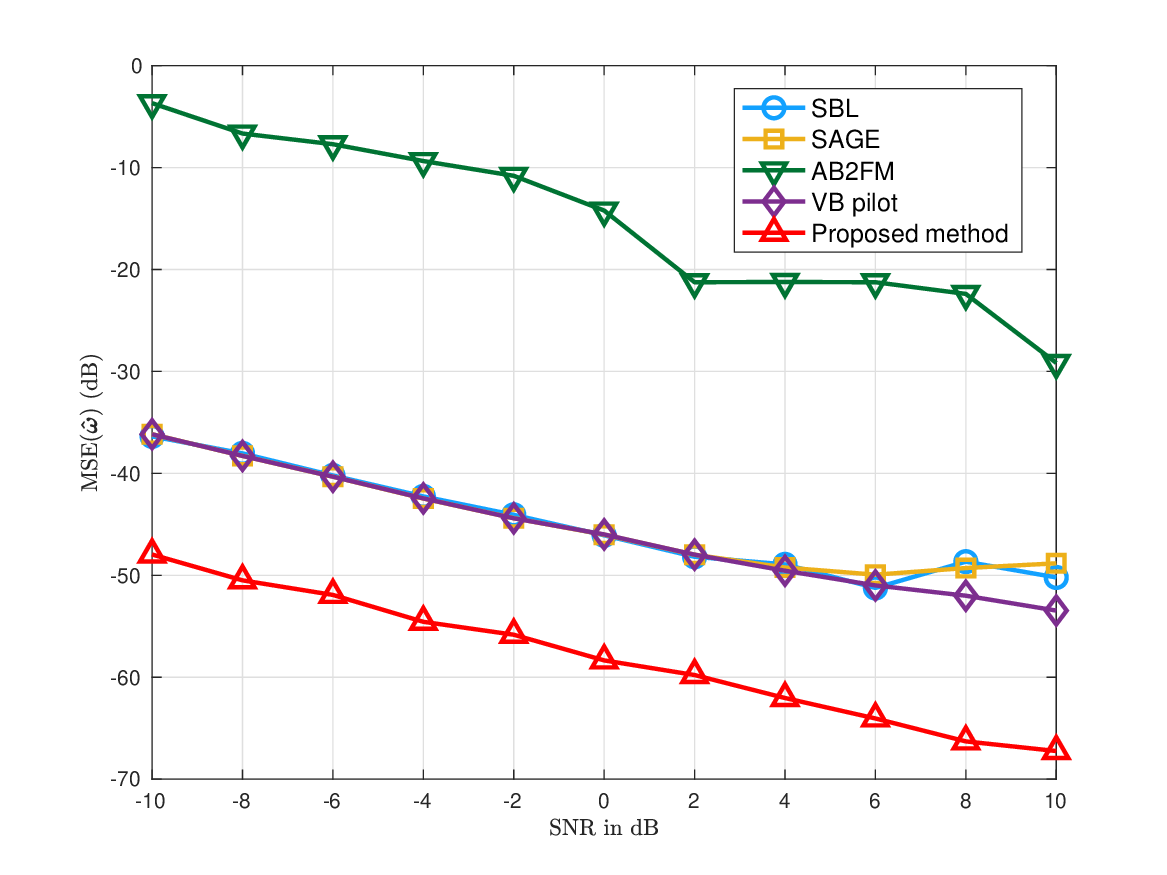}
	\caption{MSE of azimuth-angle estimation versus SNR.}
	\label{Az}
\end{figure}

\begin{figure}[!th]
	\centering
\includegraphics[width=1.0\linewidth]{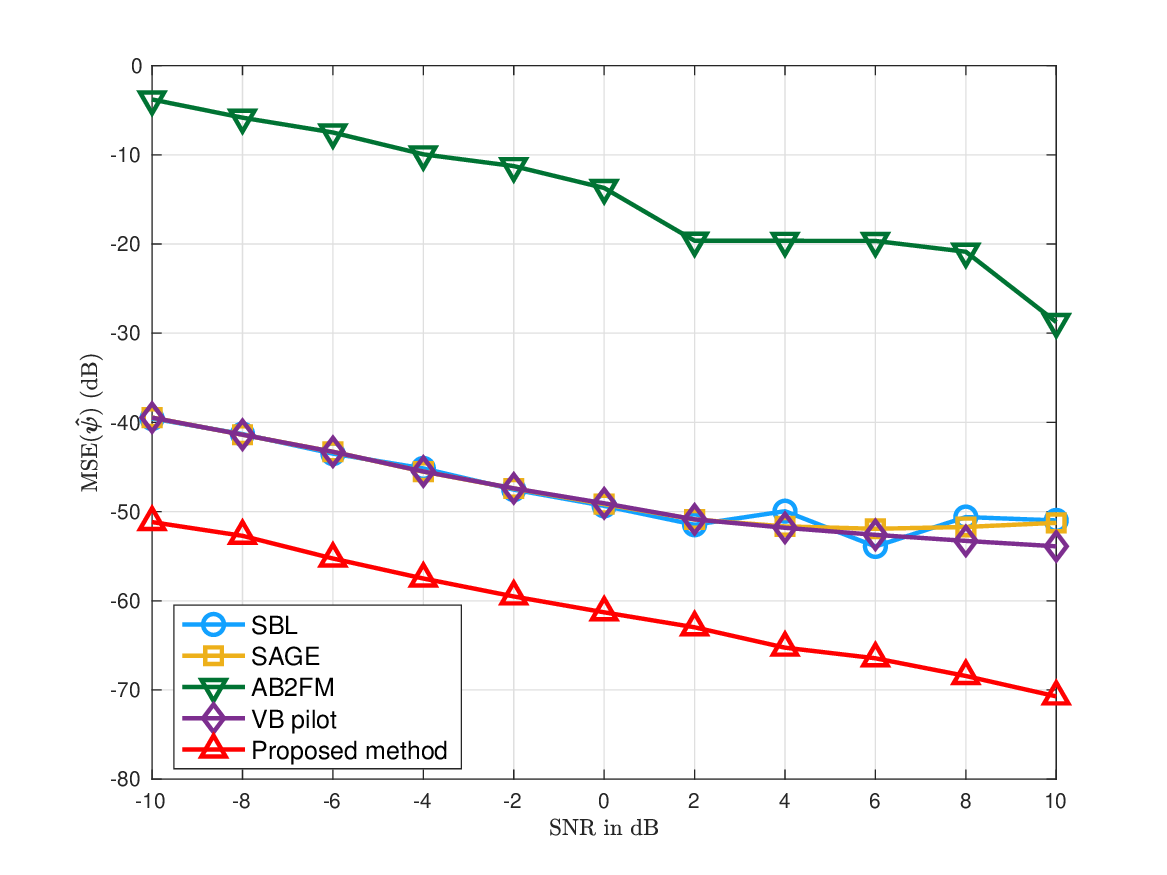}
	\caption{MSE of elevation-angle estimation versus SNR.}
	\label{ele}
\end{figure}

Figs.~\ref{Az} and~\ref{ele} show the MSE performance of azimuth- and 
elevation-angle estimation as a function of SNR. The proposed method achieves 
the best performance for both angular components over the entire SNR range. 
The pilot-only VB method also performs slightly better than SAGE and SBL, 
indicating the advantage of the proposed Bayesian inference formulation for 
gridless angular estimation. Specifically, instead of relying on an exhaustive 
search or a predefined angular dictionary, the proposed method approximates the 
angular posterior distributions by von Mises distributions, which leads to 
efficient and tractable updates. Furthermore, the data-aided refinement stage 
provides additional reliable observations through the detected data symbols, 
thereby substantially improving the accuracy of both azimuth and elevation 
estimation.

\begin{figure}[!th]
	\centering
\includegraphics[width=1.0\linewidth]{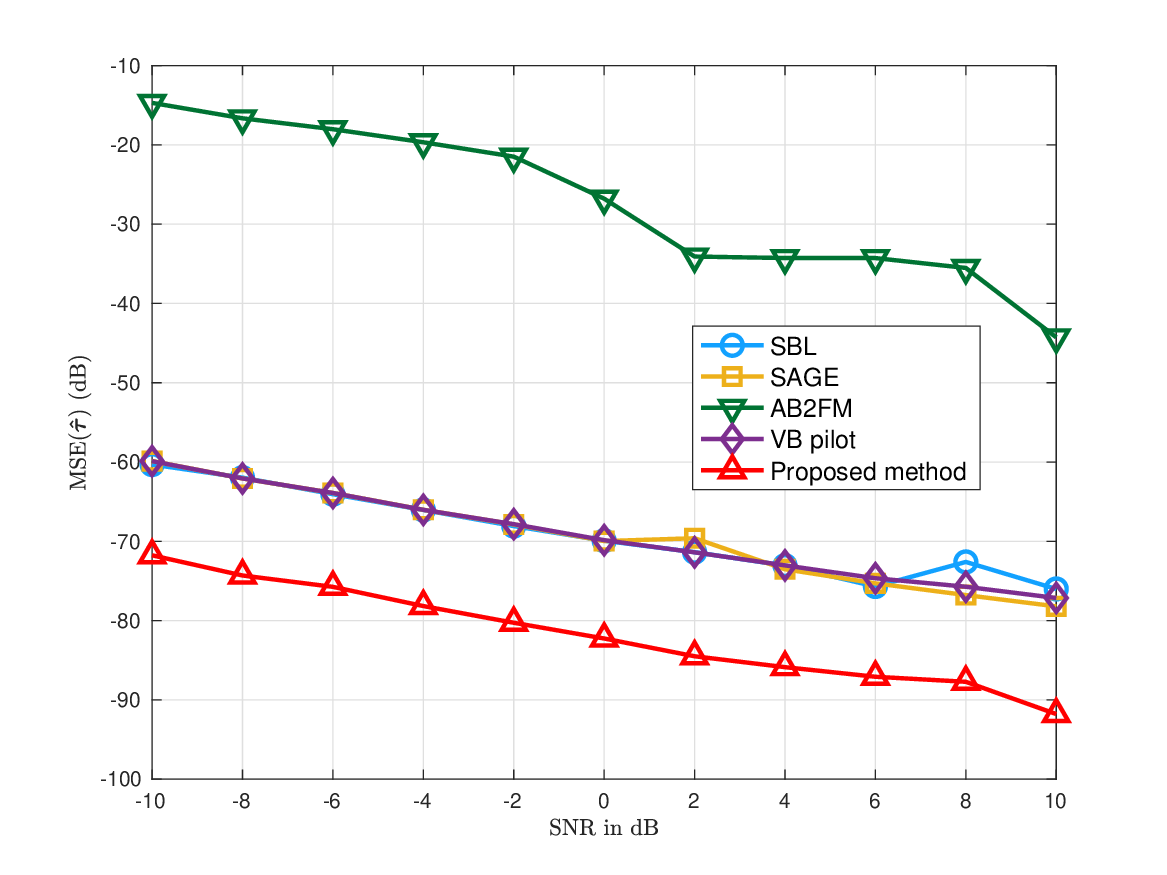}
	\caption{MSE of path-delay estimation versus SNR.}
	\label{delay}
\end{figure}

\begin{figure}[!th]
	\centering
\includegraphics[width=1.0\linewidth]{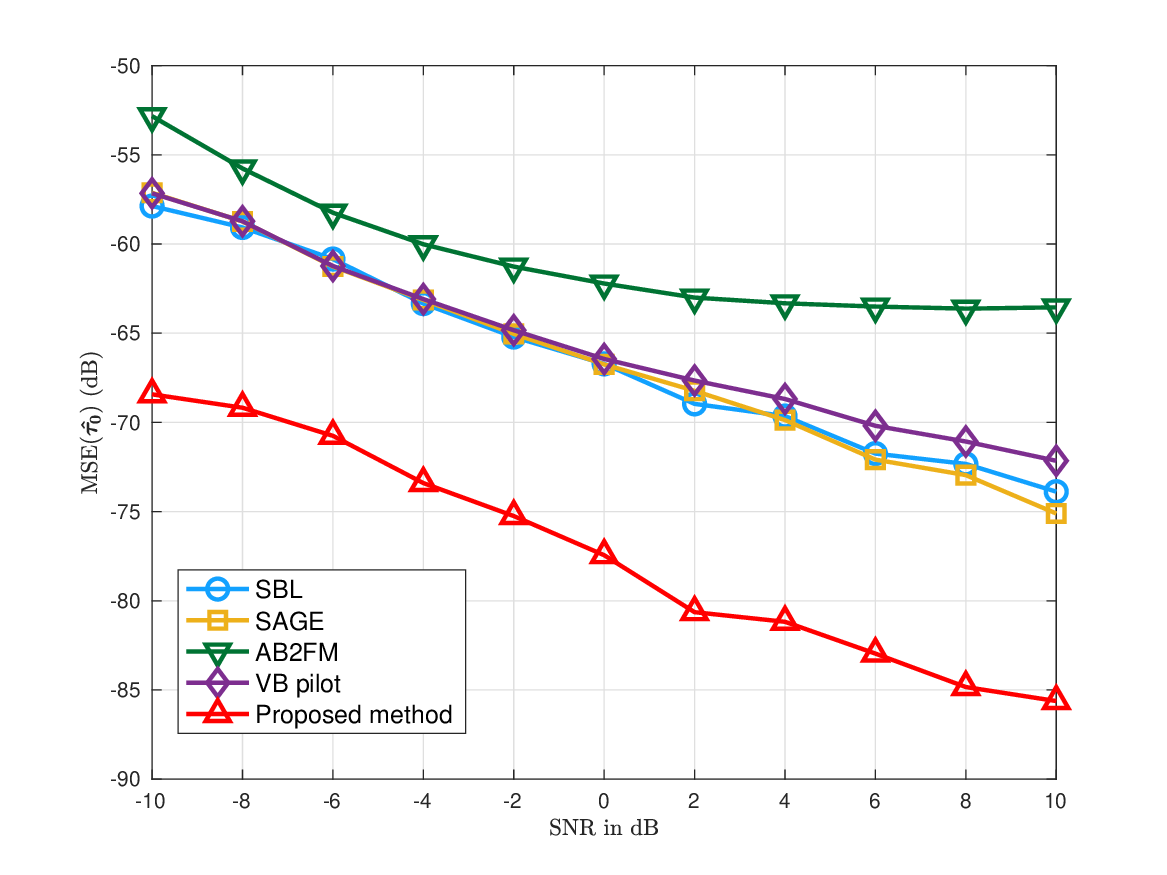}
	\caption{MSE of timing-offset estimation versus SNR.}
	\label{TO}
\end{figure}

Figs.~\ref{delay} and~\ref{TO} show the MSE performance of path-delay and 
timing-offset estimation, respectively. The proposed method consistently 
outperforms all baseline schemes across the considered SNR range. In the 
pilot-only case, the VB method achieves accuracy comparable to SAGE and SBL, 
which demonstrates the effectiveness of the proposed inference formulation for 
delay-domain parameter estimation. AB2FM performs worse than the other baselines, especially at moderate and high SNR, which suggests that it is more sensitive to residual synchronization mismatch and multipath coupling. SAGE, SBL, and pilot-only VB follow similar trends because they rely mainly on pilot observations. Their performance improves with SNR but the gain becomes limited at high SNR, likely due to limited pilot resources, grid mismatch in SBL, and local convergence sensitivity in SAGE. With the data-aided refinement stage, the 
proposed method further improves the estimation accuracy by exploiting the 
detected data symbols as additional observations. This leads to a substantial 
performance gain for both path-delay and TO estimation, confirming the advantage 
of the proposed JED-VB framework in asynchronous OFDM-ISAC systems.

\begin{figure}[!th]
	\centering
\includegraphics[width=1.0\linewidth]{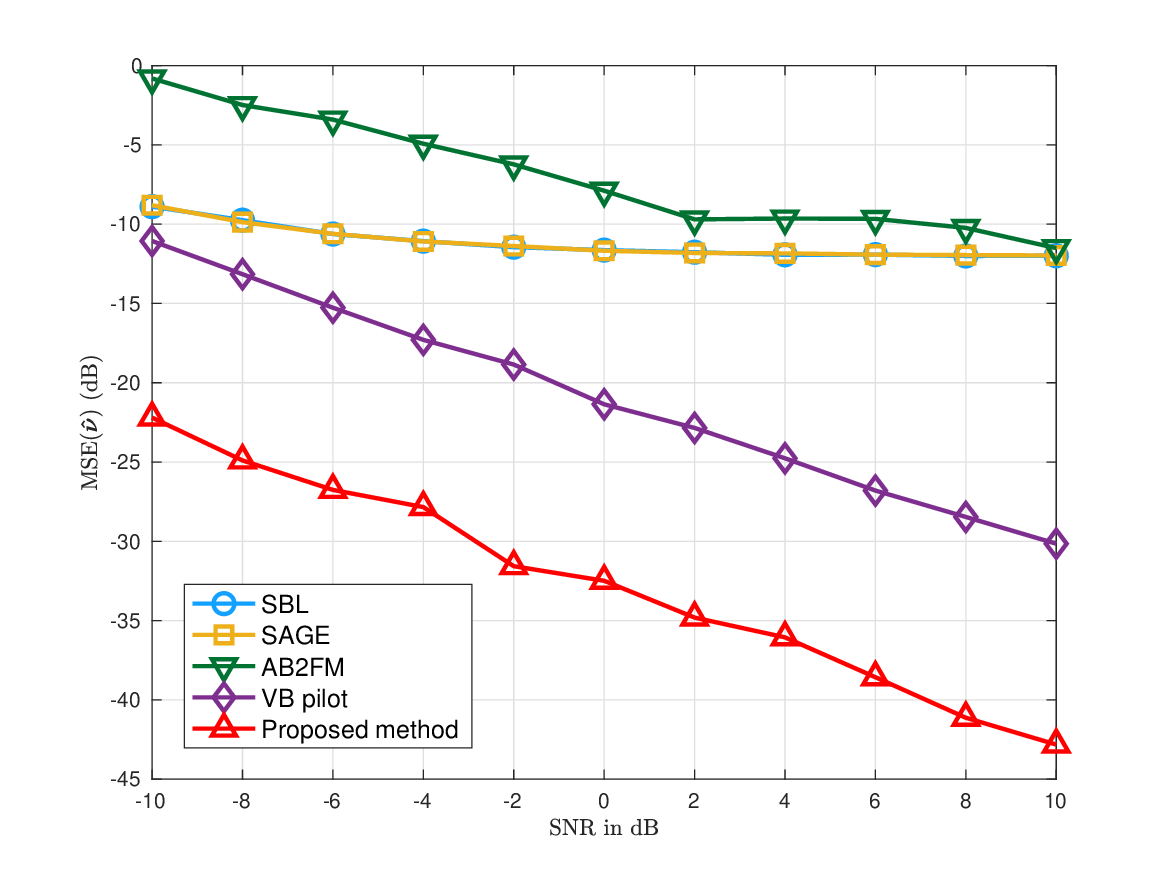}
	\caption{MSE of velocity estimation versus SNR.}
	\label{Velocity}
\end{figure}

\begin{figure}[!th]
	\centering
\includegraphics[width=1.0\linewidth]{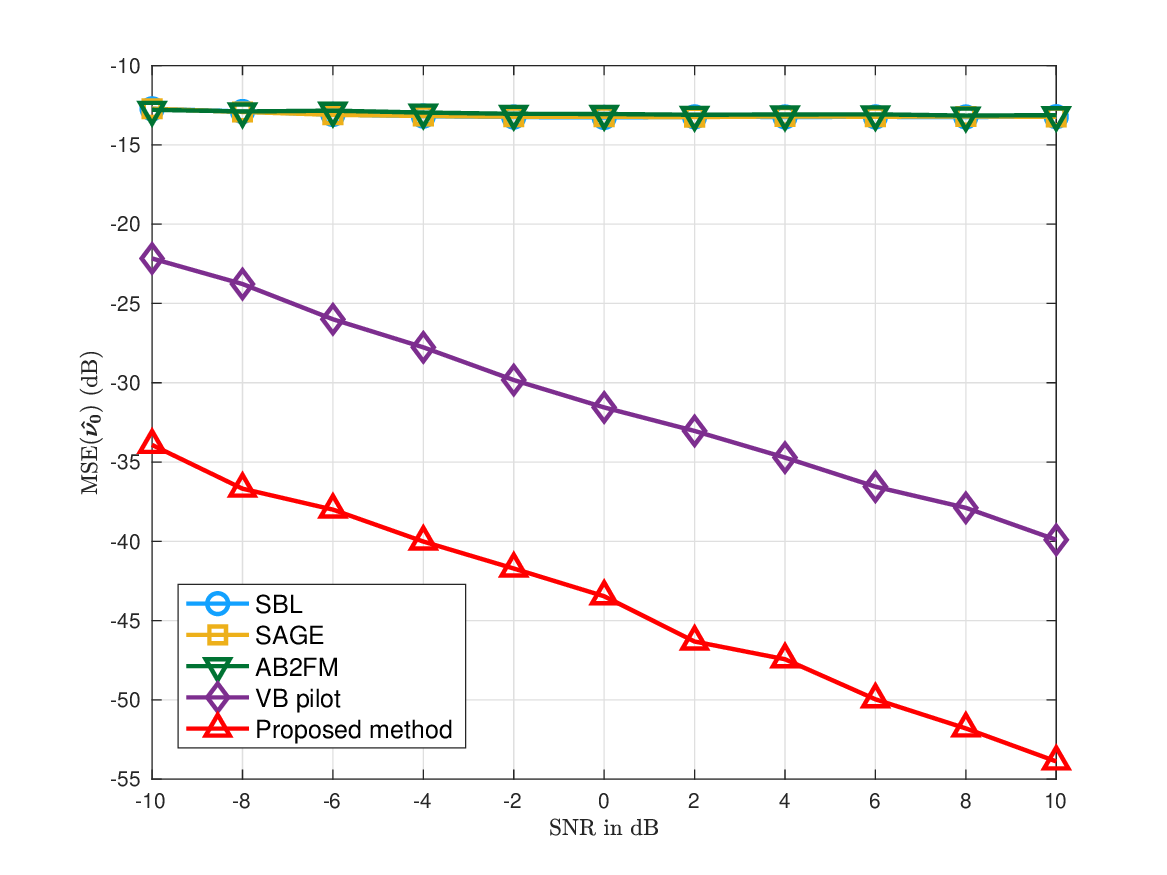}
	\caption{MSE of CFO estimation versus SNR.}
	\label{CFO_}
\end{figure}

Figs.~\ref{Velocity} and~\ref{CFO_} present the MSE performance of velocity and 
CFO estimation, respectively. The proposed method achieves the best performance 
over the entire SNR range. For velocity estimation, the pilot-only VB method 
already outperforms SAGE, SBL, and AB2FM, demonstrating the effectiveness of the 
proposed inference framework for Doppler estimation. For CFO estimation, the 
baselines exhibit clear performance saturation, while the VB-based methods 
continue to improve with SNR because the CFO is explicitly modeled and jointly 
updated with the channel parameters. With data-aided refinement, the proposed 
method further reduces the estimation error by exploiting detected data symbols 
as additional observations.

\subsection{3D Target Position Estimation}
This subsection illustrates the impact of residual TO and CFO on sensing performance in the considered OFDM-ISAC system.  Let $\mathbf p_{\rm U}\in\mathbb R^3$ and $\mathbf p_{\rm B}\in\mathbb R^3$ denote the known UE and BS positions, respectively, and let $\mathbf p\in\mathbb R^3$ denote the target position. Suppose that the BS estimates the target azimuth and elevation as $(\hat\omega,\hat\psi)$. The corresponding unit direction vector is
\begin{equation}
\hat{\mathbf u}
=
\begin{bmatrix}
\cos \hat\psi \cos \hat\omega \\
\cos \hat\psi \sin \hat\omega \\
\sin \hat\psi
\end{bmatrix}.
\end{equation}
Hence, the target lies on the ray
\begin{equation}
\mathbf p=\mathbf p_{\rm B}+r\hat{\mathbf u},
\label{eq:ray_model}
\end{equation}
where $r\ge 0$ is the BS-to-target distance. Let
\begin{equation}
d_{\rm UB}\triangleq \|\mathbf p_{\rm U}-\mathbf p_{\rm B}\|
\end{equation}
denote the UE-BS baseline. If $\hat\tau$ denotes the estimated bistatic excess delay relative to the direct UE-BS path, then
\begin{equation}
c\hat\tau
=
\|\mathbf p-\mathbf p_{\rm U}\|
+
\|\mathbf p-\mathbf p_{\rm B}\|
-
d_{\rm UB}.
\label{eq:bistatic_excess_delay}
\end{equation}
Substituting \eqref{eq:ray_model} into \eqref{eq:bistatic_excess_delay} gives
\begin{equation}
r+\|\mathbf p_{\rm B}+r\hat{\mathbf u}-\mathbf p_{\rm U}\|
=
d_{\rm UB}+c\hat\tau.
\end{equation}

Define
\begin{equation}
\mathbf b \triangleq \mathbf p_{\rm U}-\mathbf p_{\rm B},
\qquad
d \triangleq \|\mathbf b\|,
\qquad
s \triangleq d+c\hat\tau.
\end{equation}
Then
\begin{equation}
r+\|r\hat{\mathbf u}-\mathbf b\|=s.
\end{equation}
Squaring both sides and using $\|\hat{\mathbf u}\|=1$ yields
\begin{equation}
r^2-2r\hat{\mathbf u}^T\mathbf b+d^2
=
s^2-2sr+r^2,
\end{equation}
which leads to
\begin{equation}
\hat r
=
\frac{s^2-d^2}{2\left(s-\hat{\mathbf u}^T\mathbf b\right)}.
\label{eq:r_bistatic}
\end{equation}
Therefore, the 3D target position estimate is
\begin{equation}
\hat{\mathbf p}
=
\mathbf p_{\rm B}
+
\hat r\,\hat{\mathbf u}.
\label{eq:target_position_estimate}
\end{equation}

Equivalently, in Cartesian coordinates,
\begin{align}
\hat x
&=
x_{\rm B}+\hat r\cos\hat\psi\cos\hat\omega,\\
\hat y
&=
y_{\rm B}+\hat r\cos\hat\psi\sin\hat\omega,\\
\hat z
&=
z_{\rm B}+\hat r\sin\hat\psi.
\end{align}

\begin{figure}[!th]
	\centering
\includegraphics[width=1.0\linewidth]{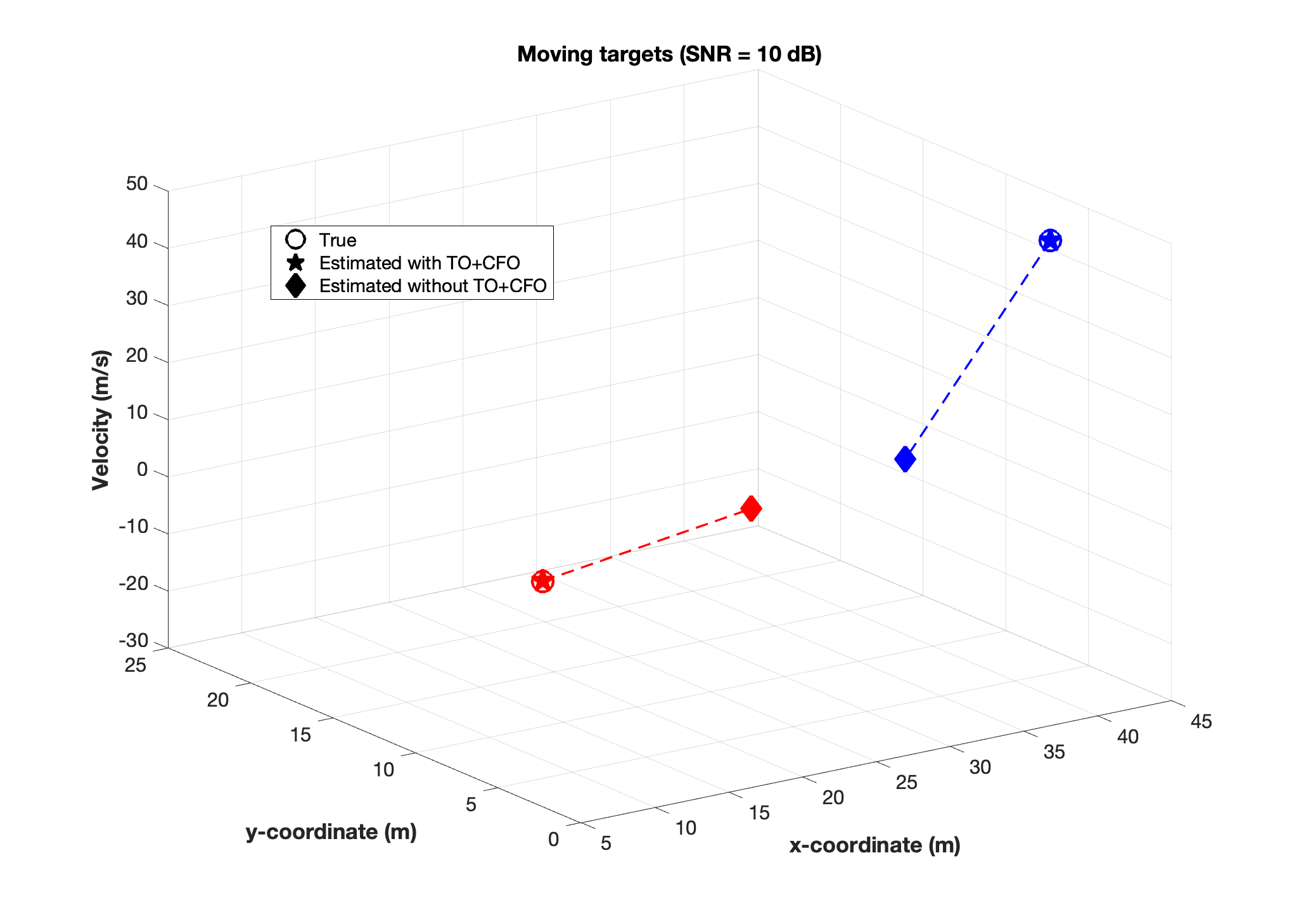}
	\caption{3D position and velocity estimation of moving targets with and without TO/CFO compensation.}
	\label{3D}
\end{figure}
Fig.~\ref{3D} shows the 3D position and velocity estimation results at 
$\mathrm{SNR}=10$ dB. With TO and CFO compensation, the proposed method 
accurately recovers the moving targets. However, when TO and CFO are ignored, 
the estimated target states deviate significantly from the true ones. This 
demonstrates that residual timing and frequency offsets can cause severe 
localization and velocity-estimation errors in OFDM-ISAC systems. Hence, it is 
important to design receivers that jointly account for synchronization, 
communication, and sensing tasks to ensure reliable data detection and accurate 
target parameter estimation.

\section{Conclusion}

In this paper, we developed a joint channel estimation, synchronization, and 
data detection framework for asynchronous uplink OFDM-ISAC systems. By explicitly 
modeling the residual TO and CFO, the proposed method jointly infers the transmitted data symbols, multipath channel parameters, and synchronization parameters within a unified VB framework. In particular, the UPA steering response was expressed in a separable form, which enables efficient 
gridless estimation of the angular parameters through von Mises distributions. Furthermore, a data-aided refinement stage was introduced to exploit the detected data symbols as additional observations, thereby improving both communication and 
sensing performance.
Simulation results demonstrated that the proposed method outperforms SBL, SAGE, 
AB2FM, and the pilot-only VB method across various metrics, including SER, channel NMSE, channel's parameters estimation, as well as TO and CFO estimation. The results further indicate that neglecting TO and CFO significantly impairs localization accuracy and velocity estimation. This underscores the need for synchronization-aware receiver designs in ISAC systems. Overall, the findings demonstrate that the proposed JED-VB framework enables robust data detection and precise sensing-parameter estimation in asynchronous OFDM-ISAC scenarios.

\bibliographystyle{IEEEtran}
\bibliography{Refs}
\end{document}